\documentclass[12pt]{article}
\pdfoutput=1
\usepackage{graphicx,amsmath,amssymb,amsfonts}
\usepackage{enumerate,setspace}
\usepackage[greek,english]{babel}
\usepackage{calligra}
\usepackage[small, bf] {caption}
\usepackage[T1]{fontenc}
\usepackage{psfrag}
\usepackage{upgreek}
\usepackage{color}   

\newcommand{\DLK}{\kappa}

\usepackage{makeidx}
\renewcommand{\a}{{\bf a}}
\usepackage{cite}
 \usepackage{hypernat}
\newcommand{\w}{{\cal W}}
\renewcommand{\b}{{\bf b}}

\newcommand{\g}{g}
\newcommand{\Tr}{E}
\newcommand{\ol}{\overline}
\newcommand{\DLc}{c}
\newcommand{\ep}{\sigma}

\renewcommand{\a}{{\bf a}}
\renewcommand{\b}{{\bf b}}
\renewcommand{\P}{{\cal P}}

\newcommand{\DLtr}{\epsilon}

\newcommand{\Mdel}{\langle\delta\rangle}
\newcommand{\eff}{\tilde}
\newcommand{\W}{{\cal W}}
\newcommand{\eq}{{\rm eq}}

\usepackage{fullpage}
\newcommand{\EQ}{\begin{equation}}
\newcommand{\EE}{\end{equation}}
\newcommand{\EQA}{\begin{eqnarray}}
\newcommand{\EEA}{\end{eqnarray}}

\def\longrightharpoonup{\relbar\joinrel\rightharpoonup}
\def\longleftharpoondown{\leftharpoondown\joinrel\relbar}
\def\longrightleftharpoons{
  \mathop{
    \vcenter{
      \hbox{
	\ooalign{
	  \raise1pt\hbox{$\longrightharpoonup\joinrel$}\crcr
	  \lower1pt\hbox{$\longleftharpoondown\joinrel$}
	}
      }
    }
  }
}

\title{Evolution of molecular phenotypes \\under stabilizing selection}
\author{Armita Nourmohammad$^{1,2,*}$, Stephan Schiffels$^{1,3,*}$, Michael L\"assig$^1$}
\date{\small $^1$ Institute f\"ur Theoretische Physik, Universit\"at zu K\"oln, Z\"ulpicherstr. 77, \\
50937 K\"oln, Germany\\
$^2$Joseph Henry Laboratories of Physics and Lewis-Sigler Institute for Integrative Genomics,\\ Princeton University,
 Princeton, NJ 08544\\
$^3$ Wellcome Trust Sanger Institute, Hinxton, Cambridge, CB10 1SA, United Kingdom\\
$^*$ Authors with equal contributions\\
}

\begin{document}

\maketitle
\begin{abstract} 
Molecular phenotypes are important links between genomic information and organismic functions, fitness, and evolution. Complex phenotypes, which are also called quantitative traits, often depend on multiple genomic loci. Their evolution builds on genome evolution in a complicated way, which involves selection, genetic drift, mutations and recombination. Here we develop a coarse-grained evolutionary statistics for phenotypes, which decouples from details of the underlying genotypes. We derive approximate evolution equations for the distribution of phenotype values within and across populations. This dynamics covers evolutionary processes at high and low recombination rates, that is, it applies to sexual and asexual populations. In a fitness landscape with a single optimal phenotype value, the phenotypic diversity within populations and the divergence between populations reach evolutionary equilibria, which describe stabilizing selection. We compute the equilibrium distributions of both quantities analytically and we show that the ratio of mean divergence and diversity depends on the strength of selection in a universal way: it is largely independent of the phenotype's genomic encoding and of the recombination rate. This establishes a new method for the inference of selection on molecular phenotypes beyond the genome level. We discuss the implications of our findings for the predictability of evolutionary processes. 
\end{abstract}

\section{Introduction}

In recent years, we have witnessed an enormous growth of information from genome sequence data, which has enabled large-scale comparative studies within and across species. How this genomic information translates into biological  functions is much less known. Molecular functions integrate the genomic information of their constitutive sites, and they can often be associated with specific phenotypes. Many such phenotypes are quantitative traits: they have a continuous spectrum of values and depend on multiple genomic sites. For example, the binding of a transcription factor to a regulatory DNA site can be monitored by its effect on the expression level of the regulated gene. 

In an evolutionary context, biological functions and their associated phenotypes are quantified by their contribution to the fitness of an organism. Such fitness effects can in principle be uncovered from genomic data by comparative analysis. However, a genome-based analysis of phenotypic evolution can be exceedingly complicated. The source of these complications is two-fold: Quantitative traits often depend on numerous and in part unknown genomic sites. Moreover, the evolution of these sites is coupled by fitness interactions (epistasis) and by genetic linkage. For both reasons, the genomic basis of a complex phenotype is not completely measurable. At the same time, details of site content, linkage, and epistasis should not matter for the evolution of the phenotype itself. This calls for an effective, coarse-grained picture of the evolutionary process at the phenotypic level, which is the topic of this paper. We will show that complex quantitative traits have  {\em universal} phenotypic observables, which decouple from the trait's genomic basis. Such universality turns out to be important for the practical analysis of a quantitative trait: it provides a way to infer its fitness effects  based solely on phenotypic measurements. 

The map from genotype to phenotype is a challenging problem for statistical theory. The reason is that epistasis and linkage generate correlations in a population: the population frequency of individuals with a combination of alleles at a set of genomic sites may be larger or smaller than the product of the single-site allele frequencies. We refer to these correlations by the standard term linkage disequilibrium (which is quite misleading, because linkage correlations have nothing to do with disequilibrium). Linkage disequilibrium is strongest in asexually evolving populations, but it can also be maintained under sexual reproduction, whenever recombination between genomic loci is too slow to randomize allele associations~\cite{Comeron:2002vn}. Linkage disequilibrium and epistasis can make the genomic evolution of a quantitative trait a strongly correlated many-``particle'' process, and these correlations are crucial for the resulting phenotype statistics. Any quantitative understanding of this dynamics must be based on an evolutionary model that contains selection, mutations, genetic drift, and (in sexual populations) recombination -- and is yet analytically tractable at least in an approximate way. Before we turn to the agenda of this paper, we briefly summarize current models of genome evolution and their application to quantitative traits. 

All known analytically solvable genome evolution models for multiple sites are based on the assumption that linkage correlations vanish~\cite{LANDE:1976tp,Falconer:1989tg,Wright:1937,Barton:1989vm,Lynch:1998vx,Johnson:2002ua,deVladar:2011bs} or are small~\cite{Neher:2011wc, BT1991}. There are two classes of quantitative traits to which these models can be applied. One of these consists of phenotypes that depend only on a small number of genomic sites. Such phenotypes are mostly monomorphic and occasionally polymorphic at a single of their constitutive sites. Hence, allele changes at different sites are well separated in time and linkage disequilibrium is small, regardless of the level of recombination. An example of such {\em microscopic} traits is transcription factor binding sites in prokaryotes and simple eukaryotes, which typically have about ten functional bases. In a time-independent fitness landscape, the  genomic and phenotypic evolution of microscopic traits leads to  simple equilibrium states of Boltzmann form, which can be used for the inference of selection (this type of equilibrium is reviewed in the next section)~\cite{Berg:2004dz,Sella:2005da}. In this way, fitness landscapes for transcription factor binding, which depend on the binding energy as molecular phenotype, have been inferred from site sequence data in bacteria and yeast~\cite{ML05,Kinney:2008tb}.

The other, complementary class is phenotypes with a large number of constitutive sites which are assumed to evolve under rapid recombination, so that linkage correlations remain small. This assumption is justified in sexual populations, if all of the sites are at sufficient sequence distance from each other. It implies that the phenotype distribution in a population is completely determined by the allele frequencies at the constitutive sites. Examples are an organism's height, complex disease phenotypes or longevity, which depend on multiple genes on different chromosomes. Such {\em macroscopic} traits are always polymorphic at multiple constitutive sites, which leads to a distribution of trait values in a population. Macroscopic traits are the traditional subject of quantitative genetics, which focuses on a phenomenological description of these trait distributions~\cite{Fisher:1930wy,LANDE:1976tp,Barton:1989vm,Falconer:1989tg,Lynch:1998vx,Rice:1990vo,Hartl:1996td,Blumer:1972,Barton:1986,Wright:1935Jg,Wright:1937}. Rapid recombination is a crucial ingredient for existing evolutionary models of macroscopic traits~\cite{Johnson:2002ua,deVladar:2011bs}. As long as linkage disequilibrium remains small, macroscopic traits also reach genomic and phenotypic Boltzmann equilibria in a time-independent fitness landscape (for details, see next section)~\cite{Fisher:1930wy,Johnson:2002ua,Vladar:2011kz,deVladar:2011bs,Wright:1937,Wright:1935}.

Many interesting molecular phenotypes, however, cannot be assumed to evolve close to linkage equilibrium. The stability of protein and RNA folds depend on their coding sequence~\cite{Smith:1970uz,Fernandez:2011uj}, protein binding affinities depend on the nucleotides encoding the binding domain~\cite{BH,Berg:2004dz}, complex regulatory interactions depend on cis-regulatory modules with several binding sites~\cite{DAVIDSON, Ptashne}, histone-DNA binding involves segments of about 150 base pairs~\cite{Radman:2010}: these are typical examples of intermediate-level phenotypes with tens to hundreds of constitutive DNA sites. Such {\em mesoscopic} phenotypes, which are often building blocks of macroscopic traits, are generically polymorphic at several constitutive sites. In asexual populations, mesoscopic traits always evolve under substantial linkage disequilibrium. This dynamics governs, for example, the evolution of antibiotic resistance in bacteria~\cite{Weinreich:2006vw} and the antigenic evolution in  human influenza A~\cite{Strelkowa}. But mesoscopic traits can build up linkage disequilibrium even in sexual populations, because their constitutive sites are localized in a small genomic region, which limits the power of recombination~\cite{Comeron:2002vn}. Genomic evolution of multiple sites under weak recombination is a strongly correlated process, which generates cooperative phenomena such as clonal interference and background selection~\cite{Charlesworth:1994vaa,Gerrish:1998wo,Barton:1995wya,Park:2007wu,Desai:2007wv,Rouzine:2008ww,Schiffels:2011fu}. In other words, the phenotype distribution in a population is no longer determined by the allele frequencies at the constitutive sites, but it depends on the full distribution of genotypes. In this volume,  Shraiman and colleagues show that the buildup of sequence correlations with decreasing recombination rate leads to a transition from allele selection to genotype selection, which is analogous to the glass transition in the thermodynamics of disordered systems~\cite{shraiman:2012}. These correlations lead to the breakdown of known analytical models for quantitative trait evolution. 
 
The evolution of molecular traits under genetic linkage is the focus of this paper. Our dynamical model for phenotypes is grounded on the evolution of their constitutive genotypes by selection, mutations, and genetic drift, which is reviewed in Section 2. In Section 3, we derive approximate, self-consistent equations for the asexual evolution of trait values in a population, which are parametrized by their mean and variance (called trait diversity). We show that this dynamics is quite different from trait evolution in sexual populations: In a time-independent fitness landscape, the joint distribution of trait mean and diversity converges to a non-equilibrium stationary state, yet the marginal distributions of both quantities still reach solvable evolutionary equilibria. In Section~4, we apply this model to evolution in a fitness landscape with a single trait optimum, where these equilibria describe the trait statistics under stabilizing selection. We compute the expected equilibrium diversity within populations, the divergence across populations, and the distance of a population from the fitness peak. In Section~5, we derive the statistics of population fitness and entropy in the equilibrium ensemble. Specifically, we compute the genetic load, which is defined as the difference between the maximum fitness and the mean population fitness, and the fitness flux, which quantifies the total amount of adaptation between the neutral state and the stabilizing-selection equilibrium. The equilibrium entropy statistics is shown to determine the predictability of evolutionary processes from single-population data. All our analytical results are confirmed by numerical simulations. 

Throughout the paper, we compare our derivations and results for non-recombining populations with their counterparts for rapid recombination. In both processes, stabilizing selection reduces trait divergence and diversity, and its effects on divergence are always stronger than on diversity. The reason will become clear in Section~4: the effective strength of selection on trait divergence is stronger than on diversity, albeit for different reasons in sexual and in asexual populations. 
In particular, we show in Section~6 that the equilibrium ratio of trait divergence and diversity shows a nearly universal behavior: it decreases with increasing strength of selection in a predictable way, but it depends only weakly on the number of constitutive sites, their selection coefficients, and the recombination rate. Hence, this ratio provides a new, quantitative test for stabilizing selection on quantitative traits, which does not require genomic data and is applicable at arbitrary levels of recombination. The agenda of the paper is summarized in Fig.~1. For the reader not interested in any technical details, the summary of genome evolution (Section~2.4) together with the basics of trait statistics (Section~3.1) and stabilizing selection (first part of Section~4) provide a fast track to the selection test in Section~6. 

\begin{figure}
\hspace{1.5cm}
\includegraphics[width=0.85\textwidth]{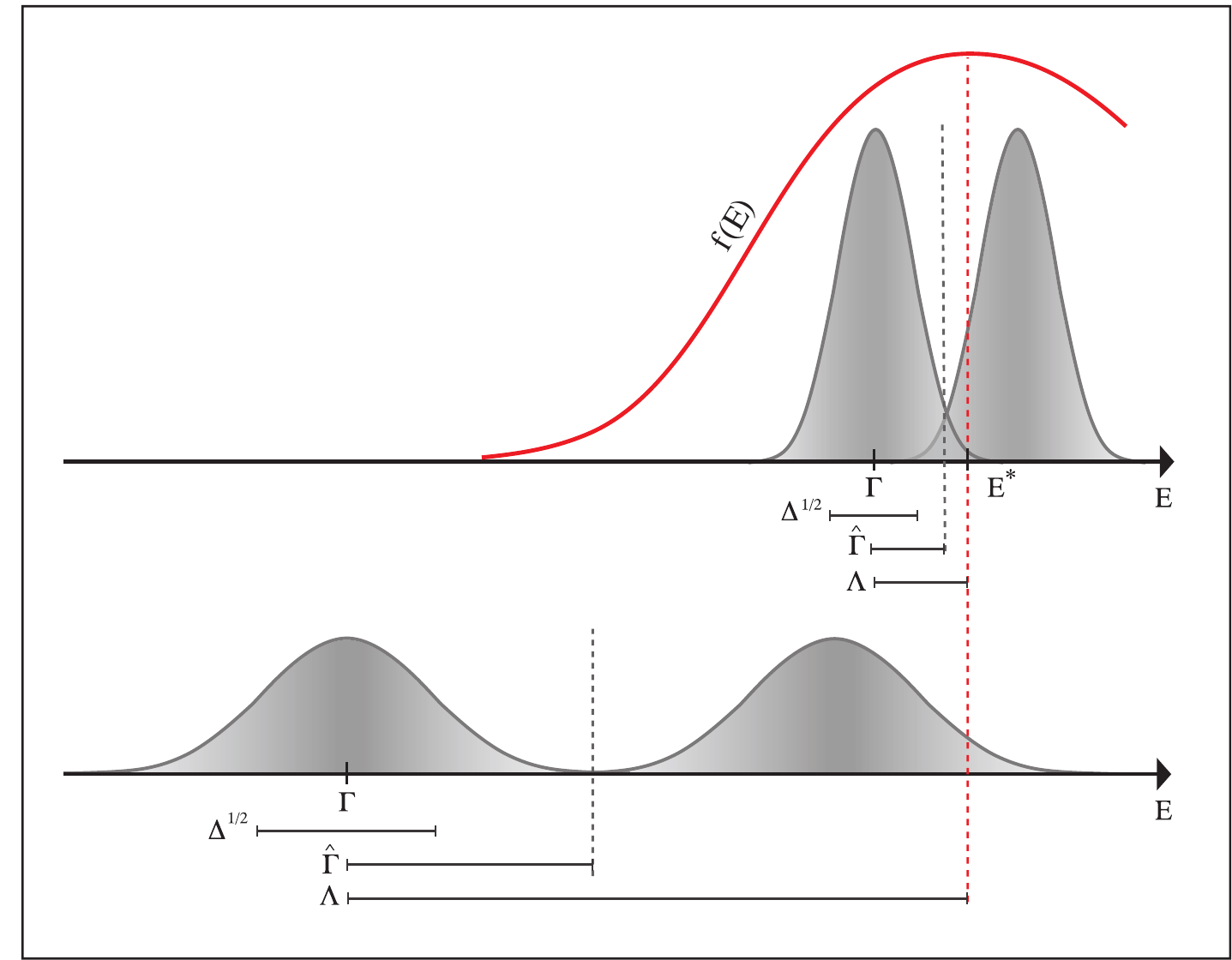}
\caption{\label{fig:schematic}
{\bf \small Evolution of a quantitative trait under stabilizing selection (schematic).} \small The trait $E$ evolves in a fitness landscape $f(E)$ favoring a single trait value $E^*$ (red line, upper panel), which can be compared to neutral evolution (lower panel). Its dynamics is a stochastic process, which results from the underlying genome evolution (Section 2). This process can be described by an ensemble of populations (Section~3). An individual population from the ensemble has a trait distribution with mean $\Gamma$ and variance (diversity) $\Delta$; two such populations are shown as brown curves. The trait mean is at a distance $\hat \Gamma \equiv \Gamma - \langle \Gamma \rangle$ from the ensemble average $\langle \Gamma \rangle$ and at a distance $\Lambda \equiv \Gamma - E^*$ from the optimal trait value $E^*$. The phenotypic population ensemble is characterized by the average divergence between populations (which equals twice the ensemble variance $ \langle \hat \Gamma^2 \rangle$), the average diversity $\langle \Delta \rangle$, and the average distance from the trait optimum, $\langle \Lambda \rangle$. Stabilizing selection reduces all of these quantities compared to neutrality, but the relative change is larger for $ \langle \hat \Gamma^2 \rangle$ and $\langle \Lambda \rangle$ than for $\langle \Delta \rangle$ (Section~4). Our theory describes a number of important characteristics of the population ensemble: genetic load, free fitness and predictability of evolution (Section~5). The ratio between divergence and diversity is the basis of a new test for stabilizing selection on quantitative traits (Section~6). 
}
\end{figure}

\section{Genome evolution} 

In this section, we review the sequence evolution models underlying our analysis of quantitative traits. All of these models are probabilistic. They describe the dynamics of an ensemble of populations, any one of which is described by the frequencies of its genotypes. The generic genotype frequency ensemble is quite intricate, because it is neither observable nor computationally accessible. However, this ensemble will serve as the basis for our theory of quantitative traits for non-recombining traits. The ensemble description simplifies in two well-known limit cases: the weak-mutation regime, where a population reduces to a single fixed genotype, and the strong-recombination regime, where genotype frequencies can be expressed by allele frequencies at individual genomic sites. 

\subsection{Evolution of genotypes}

At the most fundamental genomic level, a  population is a set of 
genotypes. A genotype is a sequence $\a = (a_1, \dots, a_\ell)$ of length $\ell$ from a $k$-letter alphabet (with $k=4$ in actual   genomes and $k=2$ in our simplified models);  
there are $K= k^\ell$ such genotypes. In a given population, each genotype has a frequency $x^\a  \geq 0$ with the constraint $\sum_\a x^\a = 1$. We describe the population
 state by recording the linearly independent frequencies $x = (x^1, \dots, x^{K-1})$ for a set $\cal A$ of 
  $K-1$ genotypes (the one remaining, arbitrarily chosen reference genotype $\a_K$ has the frequency $x^K = 1 - \sum_{\a \in {\cal A}} x^\a$). 

The evolution of genotypes is a stochastic process, which generates a probability distribution of genotype frequencies, 
$P(x,t)$. This distribution describes an ensemble of independently evolving ``replicate'' populations and follows a 
generalized Kimura diffusion equation~\cite{Kimura:1964,Ewens},  
\EQ 
\frac{\partial}{\partial t} P(x,t)  =  
\sum_{\a, \b \in {\cal A}}  \left [   \frac{1}{2N} \frac{\partial^2}{\partial x^\a \partial x^\b} \g^{\a \b} (x)
 -  \frac{\partial}{\partial x^\a} \big (m^\a (x) +  \g^{\a \b} (x) s_\b (x) \big )
 \right ] P(x,t).
\label{Ptgen}
\EE
Here and below, we adopt the convention that differential operators act on all functions to their right. The first term on the right hand side of eq.~(\ref{Ptgen}) accounts for stochastic 
changes of genotype frequencies by reproductive fluctuations in a finite population (i.e., by genetic drift). This term is proportional to the inverse of the effective population size 
$N$ and to the diffusion coefficients 
\EQ
g^{\a \b} (x)  = \left \{ 
\begin{array}{ll}
- x^\a x^\b & \mbox{if $\a \neq \b$}
\\
x^\a (1 - x^\a) & \mbox{if $\a = \b$}.
\end{array} \right. 
\label{g}
\EE
The second term describes deterministic frequency changes by mutations. In asexually reproducing populations, the coefficients $m^\a(x)$ are given in terms 
of the mutation rates $\mu_\a^\b=\mu_{\a\rightarrow\b}$ between genotypes, 
\EQ
m^\a (x) = \sum_{\b}  \big ( \mu_\b^\a \, x^\b  - \mu_\a^\b \, x^{\a} \big ); 
\label{dxdtmut}
\EE 
in sexual populations, there are additional contributions from recombination. The third term describes natural selection. Its coefficients are fitness differences, $s_\b  \equiv f(\b)  - 
f (\a_K)$, where $f(\b)$ denotes the reproduction rate (Malthusian fitness) of a genotype $\b \in {\cal A}$ and $f(\a_K)$ is the corresponding rate for the reference genotype $\a_K$. These $k-1$ selection coefficients characterize the dynamics of the linearly independent genotype frequencies $x=(x^1,\dots,x^{K-1})$.  Here we consider the simplest case, where all reproduction rates are frequency- 
and time-independent constants. In that case, the selection coefficients $s_\b$ can be written as the gradient of a scalar fitness landscape $F(x)$, 
\EQ
s_\b (x)  = \frac{\partial}{\partial x^\b} F(x), 
\label{s}
\EE
which is simply the mean population fitness, 
\EQ
F(x)  = \bar f (x) \equiv \sum_{\a} f(\a) \, x^\a 
\label{Fx}
\EE
(see ref.~\cite{Mustonen:2010ig} for a discussion of more general cases). Although the probability distribution $P(x,t)$ of genotype frequencies gives a complete description of an 
evolving population ensemble, it is not an observable quantity. Even for moderate genome length $\ell$, there are vastly more possible genotype distributions $x$ than can be 
recorded from the history of a single population or even from an ensemble of independently evolving populations. Like the probability distribution over phase space in statistical 
mechanics, this distribution should be regarded as a conceptual and computational intermediate: $P(x,t)$ is calculated using maximum-entropy postulates, and it is used to 
define and predict expectation values of observable quantities. 

Importantly, the definition of expectation values involves averaging at two distinct levels. In a given population, the genotype frequencies $x$ determine the allele frequencies at 
individual genomic sites, 
\EQ
y_i^a \equiv \ol{\sigma_i^a} = \sum_\a \sigma_i^a \, x^\a,
\label{y}
\EE
the haplotype (allele combination) frequencies at pairs of sites,
\EQ
y_{ij}^{ab} \equiv \ol{\sigma_i^a \sigma_j^b} = \sum_\a \sigma_i^a \, \sigma_j^b \, x^\a,
\EE
and so on, which are conveniently represented as averages of the ``spin'' variables 
\EQ
\sigma_i^a \equiv \left \{ 
\begin{array}{ll}
1 & \mbox{ if  $a_i = a$,}
\\
0 & \mbox{ otherwise.}
\end{array} \right.
\label{sigma}
\EE
These averages within a population are denoted by overbars. Connected correlation functions at a single site,
\EQ
\pi_i^a \equiv \ol{(\ep_i^a - y_i^a)(\ep^a_i- y_i^a)}  = y_i^a (1  - y_i^a),
\label{pi}
\EE
are components of the sequence diversity $\pi_i = \sum_{a = 1}^{k-1} \pi_i^a$, correlations between different sites,
\EQ
\pi_{ij}^{ab} \equiv \ol{(\ep_i^a - y_i^a)(\ep^b_j - y_j^b)}  = y_{ij}^{ab} - y_i^a y_j^b
\hspace{1cm} (i \neq j), 
\label{LD}
\EE
measure linkage disequilibrium, i.e., biases in the association of alleles to haplotypes within a population.  For all of these quantities, the genotype frequency distribution $P(x,t)$ 
defines expectation values in an ensemble of independently evolving populations,
\EQ
\big \langle \, \ol{\sigma_i^a \sigma_j^b \dots} \, \big \rangle \equiv \int  \ol{\sigma_i^a \sigma_j^b \dots} \; P(x,t) \, dx;
\label{corr}
\EE
averages across populations are denoted by angular brackets, $\langle \cdot\rangle$. Such nested correlation functions can often be decomposed into independent fluctuation components within and 
across populations; for example, 
\EQ
\big \langle \, \ol{\sigma_i^a \sigma_j^b} \, \big \rangle = 
\big \langle \, \ol{\sigma_i^{a\textcolor{white}{b}}} \; \ol{\sigma_j^{b\textcolor{white}{b}}} \, \big \rangle 
+ \big \langle \, \ol{(\sigma_i^a - y_i^a) \;(\sigma_j^b - y_j^b)} \, \big \rangle = 
\big \langle y_i^a y_j^b \big \rangle + \big \langle \pi_{ij}^{ab} \big \rangle.
\EE
In particular, fitness interactions (epistasis) can generate allele frequency correlations $\big \langle y_i^a y_j^b \big \rangle$ even if linkage disequilibrium vanishes. 

All frequency correlation functions of the form (\ref{corr}) can, in principle, be computed from the solution of the diffusion equation (\ref{Ptgen}). This is impossible in practice, 
however, because no general analytical solution exists. In particular, the distribution $P(x,t)$ does not converge to an evolutionary equilibrium, which is defined as a state with 
detailed balance (see ref.~\cite{Mustonen:2009vu} for a review of detailed balance in an evolutionary context). 
We remind the reader that a diffusion equation of the form (\ref{Ptgen}) has an equilibrium distribution if and only if the vector field 
$v^\a (x) \equiv \sum_{\b \in {\cal A}} \big ( m^\a (x) +  \g^{\a \b} (x) s_\b (x) \big )$ satisfies the integrability conditions, 
\EQ
\frac{\partial}{\partial x^\b} \sum_{\a'} g_{\a \a'} (x) v^{\a'} (x) - 
\frac{\partial}{\partial x^\a} \sum_{\a'} g_{\b \a'} (x) v^{\a'}(x)
= 0,
\label{integrability}
\EE
which implies that $v^\alpha (x)$ can be written as the gradient of a scalar function. It is easy to see that the frequency-dependence of the diffusion matrix $g^{\a \b} (x)$ makes already the mutation vector field $m^\a (x)$ non-integrable. 
Hence, even if the selection coefficients $s_\b (x)$ are the gradient of a scalar fitness landscape as given by eq.~(\ref{s}), there is no general evolutionary equilibrium. We now discuss the two known special cases in which the diffusion equation (\ref{Ptgen}) does have a solvable equilibrium, which is the analogue of the Boltzmann equilibrium in statistical thermodynamics.

\subsection{Weak-mutation regime}

This regime is defined by a low genome- and population-wide mutation rate per generation, $\mu N \ell \ll 1$~\cite{Gillespie91}.  With typical values $\mu N \sim 10^{-2}$, this regime applies to short sequence  segments with a length up 
to about 10 base pairs, which are the genomic basis of {\em microscopic} traits. Transcription factor binding sites in prokaryotes and simple eukaryotes with a typical length of about 10 base pairs are examples of this kind~\cite{ML05,Kinney:2008tb}.
 In such segments, a single genotype is fixed in the population at most times. This genotype evolves through occasional polymorphisms at a single genomic site, but co-occurrence of polymorphisms at multiple sites can be neglected. We can then project the Kimura equation 
(\ref{Ptgen}) onto a Master equation on the space of fixed genotypes,
\EQ 
\frac{\partial}{\partial t} P(\a,t)  =  \sum_\b \, [u_{\b \to \a} P(\b,t) - u_{\a \to \b} P(\a,t)].
\label{Pat}
\EE
The substitution rates  are given by the classic Kimura-Ohta formula~\cite{MotooKimura:1969va,Kimura:1962um}, 
\EQ
u_{\a \to \b} = \mu_{\a \to \b} 2 N (f(\b) - f(\a)) / (1 - \exp[2 N (f(\b) - f(\a))],
\EE
with selection coefficients given by the discrete fitness landscape $f(\a)$. We make an assumption on 
neutral evolution: it occurs by point mutations with site-independent rates $\mu_{a \to b}$, which satisfy the detailed-balance relations 
\EQ
p_0 (a) \, \mu_{a \to b} = p_0 (b) \, \mu_{b \to a}. 
\label{det_bal}
\EE 
This detailed-balance assumption, which is part of all standard neutral mutation models, reduces the number of independent rate constants from 12 to 9. The resulting equilibrium single-nucleotide distribution $p(a)$ ($a = 1, \dots, k$) describes the effect of mutational biases (if all rates are symmetric, $\mu_{a \to b} = \mu_{b \to a}$, it leads to a flat single-nucleotide equilibrium $p_0 (a) = 1/k$). The detailed-balance condition in the weak-mutation regime is much weaker than the corresponding condition (\ref{integrability}) for frequency evolution, which constrains an entire function $v^\a (x)$. 

In an arbitrary fitness landscape $f(\a)$, the full dynamics (\ref{Pat}) with (\ref{det_bal}) has an equilibrium probability distribution of fixed genotypes~\cite{Berg:2004dz,Sella:2005da}
\EQ
P_\eq (\a) = \frac{1}{Z} \,  P_0 (\a) \exp [2N f(\a)],
\label{Pa_eq}
\EE   
which is the product of the factorizable neutral equilibrium 
\EQ
P_0 (\a) = \prod_{i=1}^\ell p_0 (a_i) 
\EE
and the Boltzmann factor $\exp[2N f(\a)]$. Here and below, $Z$ denotes a normalization factor. A generic fitness landscape generates cross-population allele correlations $ \big \langle y_i^a y_j^b \big \rangle$ between sites, but linkage disequilibrium vanishes without any assumptions on the recombination rate.

\subsection{Strong-recombination regime}

In this regime, linkage correlations become small because of rapid allelic reassortments in the population.
We can then approximate the frequency of a genotype by the product of its allele frequencies, 
\EQ
x^\a = y^{a_1}_1 \dots y^{a_\ell}_\ell.
\label{xLE}
\EE
This approximation, which we will refer to as {\em free recombination}, describes complete linkage equilibrium. It becomes exact in the limit of infinite recombination rate and infinite population size. Linkage equilibrium is the standard assumption of quantitative genetics~\cite{Johnson:2002ua,Fisher:1930wy,LANDE:1976tp,Falconer:1989tg,Lynch:1998vx}; it is often applied to large genomes in sexually reproducing populations, 
which are the genomic basis of macroscopic quantitative traits. Given the factorization (\ref{xLE}), we can project the Kimura equation (\ref{Ptgen}) 
onto a diffusion equation for the joint distribution of allele frequencies. In the simplest case of a two-letter genomic alphabet, this equation takes the form 
\EQ
\frac{\partial}{\partial t} P(y, t) = \sum_{i=1}^\ell \left [
\frac{1}{2N} \frac{\partial^2}{\partial y_i^2} g (y_i) - \frac{\partial}{\partial y_i} \big (m (y_i) + g (y_i) s_i (y) \big ) \right ] P(y,t).  
\label{Pyt}
\EE
Here, $y = (y_1, \dots, y_\ell)$ denotes the set of allele frequencies, 
$g(y) = y (1 - y)$ and $m(y) = \mu (1 - 2y)$ are the diffusion and mutation coefficients, and $s_i (y) = \partial F(y) / \partial y_i$ are the selection coefficients for alleles, with $F
(y) = \sum_\a f(\a) \, y^{a_1}_1 \dots y^{a_\ell}_\ell$.   In an arbitrary fitness landscape $F(y)$, the projected Kimura equation has an equilibrium distribution~\cite{Wright:1931,Wright:1935},
\EQ
P_\eq (y) = \frac{1}{Z} \,  P_0 (y) \exp[2 N F(y)], 
\label{Py_eq}
\EE
which is the product of the factorizable neutral equilibrium 
\EQ
P_0 (y) = \frac{1}{Z_0} \prod_{i=1}^\ell \; [y_i (1 - y_i)]^{-1 +2 \mu N}
\label{Py_0}
\EE
and the Boltzmann factor, $\exp[2 N F(y)]$. In this case, equilibrium emerges because the neutral distribution is the product of one-dimensional allele frequency distributions, for which the integrability condition (\ref{integrability}) is always fulfilled. Just as in the weak-mutation regime, a generic fitness landscape generates allele frequency correlations $ \big \langle y_i^a y_j^b \big 
\rangle$, which are compatible with linkage equilibrium. The strong-recombination calculus can be extended to populations with a large but finite recombination rate~\cite
{deVladar:2011bs, Neher:2011wc}. Such populations still reach an evolutionary equilibrium of the form (\ref{Py_eq}); however, the neutral distribution $P_0 (y)$ no longer factorizes and 
there are small but systematic linkage correlations $\langle \pi_{ij} \rangle$~\cite{Neher:2011wc}.

\subsection{Summary}

Genomic evolution under mutations, recombination, genetic drift, and selection can be described by a Kimura diffusion equation at the genotype level~\cite{Kimura:1964,Ewens}. The general Kimura equation does not have a closed solution. In the regimes of low mutation rate or high recombination rate, where linkage disequilibrium is small, we can project this equation onto fixed genotypes or allele frequencies, respectively. These projections are shown in Table~1. The projected equations have solvable equilibria of the form $P = P_0 \exp[2 N F]$; see equations (\ref{Pa_eq}) and (\ref{Py_eq}). The ``Boltzmann'' factor $\exp[2 N F]$ links the equilibrium probability distribution under time-independent selection, $P$, with the corresponding distribution for neutral evolution, $P_0$; this relation can serve as a starting point for the inference of selection. However, genomic equilibria do not exist for strongly coupled multi-site evolution with large linkage disequilibrium, which is common in asexual populations and  even in sexual populations for compact, intermediate-size genomic regions~\cite{Comeron:2002vn}.  

\begin{table}[h!]
\includegraphics[width=0.55\textwidth]{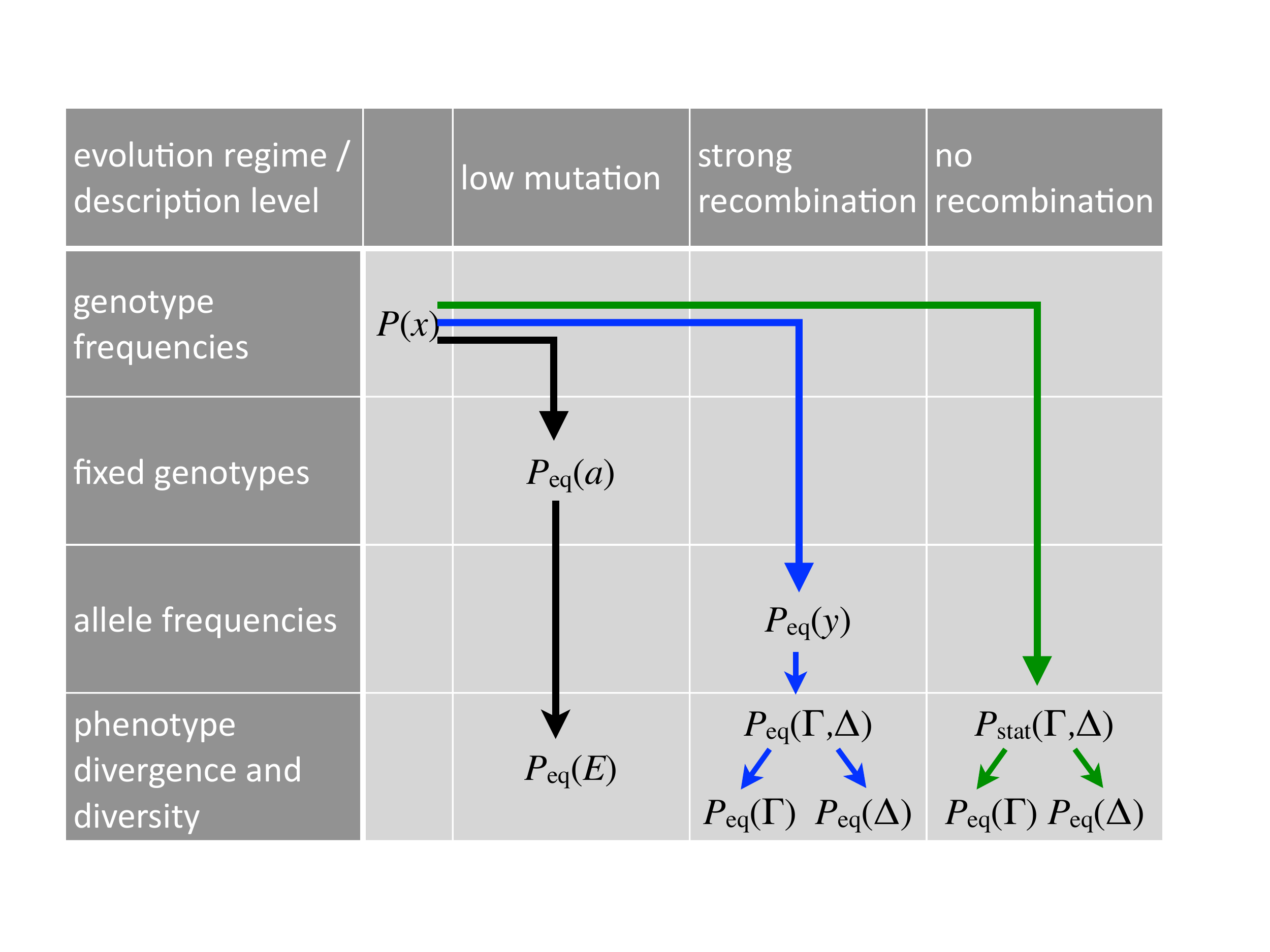}
\caption{\label{fig:projections}
{\bf From genotypes to phenotypes. } This table shows the genomic and phenotypic stationary population ensembles discussed in the text. These ensembles are obtained by different projections of the diffusive genotype dynamics (\ref{Ptgen}) under time-independent selection, which are marked by arrows. 
In the low-mutation regime, we obtain an equilibrium distribution of fixed genotypes, which can be projected further onto an equilibrium of fixed trait values; see equations (\ref{Pa_eq}) and (\ref{QE_eq}). 
In the strong-recombination regime, we obtain an equilibrium distribution of allele frequencies, which can be projected further onto a joint equilibrium of trait mean and diversity; see equations (\ref{Py_eq}) and (\ref{QGammaDelta_eq}). 
For complex traits evolving without recombination, we obtain a stationary non-equilibrium distribution of trait mean and diversity, which can be projected further onto equilibrium marginal distributions; see equations (\ref{Qstat}), (\ref{QG_eq}) and (\ref{QD_eq}). 
}
\end{table}

\section{Evolution of quantitative traits}

In this Section, we first introduce the basic statistical observables for quantitative traits. In the low-mutation and in the strong-recombination regime, we obtain phenotypic equilibria by projection from the genomic equilibrium distributions discussed in the previous Section. For complex traits evolving under linkage disequilibrium, we show that projection of the general genotype dynamics leads to a stationary non-equilibrium distribution, which describes the statistics of trait divergence and diversity under time-independent selection. Further projections lead to equilibrium marginal distributions of trait divergence and diversity, which are the basis for our subsequent analysis of stabilizing selection. The projections from genomic to phenotypic distributions are also shown in Table~1.

\subsection{Trait statistics within and across populations}

The subject of this paper is the evolution of quantitative traits with a heritable component, which depends on an individual's genotype. Here we study the simplest case of an additive map from genotype to phenotype, and we assume a binary genomic alphabet (extension to a $k$-letter alphabet is straightforward). Any phenotype $E$ can then be written in the form 
\EQ
E(\a) =  \sum_{i=1}^\ell \Tr_i \, \sigma_i 
\hspace{1cm} \mbox{with } 
\sigma_i \equiv \left \{ 
\begin{array}{ll}
1 & \mbox{ if  $a_i = a_i^*$,}
\\
0 & \mbox{ otherwise.}
\end{array} \right.
\label{EofA}
\EE
Here the phenotype is measured from its minimum value, $\a = (a_1, \dots, a_\ell)$ is the genomic sequence at its constituent sites, $a_i^*$ is the allele conferring the larger phenotype at a given site, 
 and $\Tr_i > 0$ is the phenotypic effect at that site, i.e., the difference in trait value between the two alleles. We define the allelic average $\Gamma_0$ and the overall effect amplitude $E_0$ by 
\EQ
\Gamma_0 \equiv \frac{1}{2} \sum_{i=1}^\ell E_i, 
\hspace{1cm}
E_0^2 \equiv\frac{1}{4} \sum_{i=1}^\ell E_i^2. 
\label{E0}
\EE

We are interested in the evolution of complex molecular traits, which depend on multiple genomic sites. If the number of constituent sites is sufficiently high (such that $\mu N \ell$ is of order 1 or larger), such traits are generically polymorphic in a population, even if most individual sites are monomorphic (i.e., $\theta \equiv \mu N \ll 1$, which is the case in most populations). The trait values in the individuals of a population follow a distribution $\w(E)$.  Here we parametrize this distribution by its mean and its variance, which is called the trait {\em diversity}~\cite{LANDE:1976tp,deVladar:2011bs,Buerger:1991,Barton:2009genetics}:
\EQ
\Gamma \equiv  \ol E = \sum_\a E(\a) \, x^\a, 
\hspace{1cm}
\Delta  \equiv  \ol{\Tr^2} - \Gamma^2  =  \sum_\a E_\a^2 \, x^\a -\sum_{\a,\b} E_\a E_\b x^\a x^\b
\EE
Using eq.~(\ref{y}), the trait mean can be written as a function of the allele frequencies, 
\EQ
\Gamma (y) = \sum_{i=1}^\ell  \Tr_i \ol{\ep}_i = \sum_{i=1}^\ell  \Tr_i y_i
\label{TraitAv.}
\EE
The trait diversity can be decomposed into the {\em additive trait diversity} $\Delta_1(y)$, which depends only on the allele frequencies, and the {\em trait autocorrelation} $\Delta_2(\pi)$, which depends on the linkage disequilibria between the constituent loci, 
\EQA
\Delta (y, \pi) 
&  =& \sum_{i,j =1}^\ell E_i E_j (\ol{\sigma_i \sigma_j} - \ol \sigma_i \ol \sigma_j )
 \nonumber \\
& = & \sum_{i=1}^\ell E_i^2 y_i (1 - y_i) + \sum_{i \neq j} E_i E_j \pi_{ij}  \equiv  \Delta_1 (y) + \Delta_2 ({\bf \pi}), 
\label{TraitVar.}
\EEA
where we have used eqs.~(\ref{pi}) and (\ref{LD}). As will become clear, the trait diversity is therefore more strongly affected by linkage and recombination than the trait mean. In the strong-recombination approximation of quantitative genetics, the population is at linkage equilibrium and the trait diversity reduces to its additive part, $\Delta \simeq \Delta_1$. Under finite recombination, stabilizing selection generates a negative trait autocorrelation $\Delta_2$, and the assumption of linkage equilibrium will lead to an overestimation of $\Delta$. 

Similarly to genotype evolution, the stochastic evolution of a quantitative trait generates a probability distribution $Q(\Gamma, \Delta, t)$, which describes an ensemble of independently evolving populations, each having a trait  distribution with mean $\Gamma$ and variance $\Delta$; see also~\cite{Buerger:1991}. The probability $Q(\Gamma, \Delta, t)$ is a sum of probabilities of genotype frequencies, 
\EQ
Q(\Gamma, \Delta, t) = \int \updelta(\Gamma (y(x)) - \Gamma) \, \updelta(\Delta (y(x), \pi(x))  - \Delta) \, P(x,t) dx,
\label{QGammaDelta}
\EE
where $\updelta(\cdot)$ is the Dirac delta function. Furthermore, we assume that selection acts on a trait's constituent genotypes only via the trait itself,
\EQ
f(\a) = f(E(\a)); 
\EE
that is, all genotypes with the same trait value $E$ have the same fitness $f(E)$. The genotypic fitness landscape $F(x)$ given by eq.~(\ref{Fx}) then  defines a phenotypic fitness landscape
\EQ
F(\Gamma, \Delta) \equiv \bar f (\Gamma, \Delta) = f(\Gamma) + \frac{1}{2} \Delta \, f'' (\Gamma),
\label{FGammaDelta}
\EE
which contains the leading terms in the Taylor expansion of $f(E)$ around the trait mean $\Gamma$. The phenotypic evolutionary scenario is illustrated in Fig.~1 for evolution in a single-peak landscape $f(E)$ and for neutral evolution (these cases are analyzed in detail in the next section). Each population drawn from the ensemble distribution $Q(\Gamma, \Delta)$  has a trait distribution with mean $\Gamma$ and variance $\Delta$. The trait mean of a given population is at a distance 
\EQ
\hat \Gamma \equiv \Gamma - \langle \Gamma \rangle
\EE
from the ensemble average $\langle \Gamma \rangle$, 
at a distance 
\EQ
\Lambda \equiv \Gamma - E^*
\EE
from the optimal trait value $E^*$, and two populations have a square trait distance  $(\Gamma_1 - \Gamma_2)^2$, which is called their trait  {\em divergence}. Statistical theory predicts ensemble averages such as $\langle \Gamma^2 \rangle$, $\langle \Delta \rangle$, $\langle \Lambda \rangle$, and $\langle (\Gamma_1 - \Gamma_2)^2 \rangle = 2 \langle\hat \Gamma^2 \rangle$. In this paper, we focus on stationary ensembles under time-independent selection. The trait divergence can also be defined for a single population at two different times, $D(t_2 - t_1) \equiv (\Gamma(t_2) - \Gamma(t_1))^2 $, or more generally for two populations with a common ancestor. Under time-independent selection, $D(t)$ reaches  the equilibrium ensemble divergence for long times, $\lim_{t \to \infty} \langle D(t) \rangle =  \langle (\Gamma_1 - \Gamma_2)^2 \rangle$. The statistics of time-dependent trait divergence will be analyzed in another paper\cite{HeldNourmohammadLassig12}.

The projection from genotypes to phenotypes given by eqs.~(\ref{QGammaDelta}) and (\ref{FGammaDelta}) can immediately be put to use in the regimes of low linkage disequilibrium discussed in the previous section, where an evolutionary equilibrium exists at the genomic level. In the weak-mutation regime, we obtain an equilibrium distribution of fixed phenotype values by projection  from eq.~(\ref{Pa_eq}),
\EQ
Q_\eq (E) = \frac{1}{Z} \,  Q_0(E) \exp[2 N  f(E)], 
\label{QE_eq}
\EE
this type of equilibrium distribution has been used in refs.~\cite{Berg:2004dz, ML05, Kinney:2008tb}. In the strong-recombination regime, the phenotypic equilibrium obtained by projection from eq.~(\ref{Py_eq}),
\EQ
Q_\eq (\Gamma, \Delta) =  \frac{1}{Z} \,  Q_0 (\Gamma, \Delta) \exp[2 N F(\Gamma, \Delta)]. 
\label{QGammaDelta_eq}
\EE
has been analyzed in detail in refs.~\cite{deVladar:2011bs,Barton:2009genetics}.

\subsection{Joint evolution of trait mean and diversity}

As discussed in the previous section, this equilibrium calculus is not applicable to correlated evolutionary processes in non-recombining or slowly recombining genomes, which evolve large values of linkage disequilibrium. To analyze such processes, we proceed differently: we directly use the Kimura equation for genotypes to obtain by projection a self-consistent, approximate diffusion equation for the phenotypic ensemble distribution $Q(\Gamma, \Delta, t)$. In this paper, we study the case of strictly asexual, non-recombining populations. By projection from eq.~(\ref{Ptgen}), we find the phenotypic diffusion equation
\EQA
\frac{\partial}{\partial t} Q(\Gamma, \Delta, t) & = &  
\left [ \frac{1}{2N} \left ( \frac{\partial^2}{\partial \Gamma^2} g^{\Gamma \Gamma} 
 + \frac{\partial^2}{\partial \Delta ^2} g^{\Delta \Delta} \right )
 -  \frac{\partial}{\partial \Gamma} \left ( m^\Gamma + g^{\Gamma \Gamma} s_\Gamma \right ) \right.
 \nonumber \\
 & & 
\left.  -  \frac{\partial}{\partial \Delta} \left ( m^\Delta + g^{\Delta\Delta} s_\Delta \right )
\right ] Q(\Gamma, \Delta, t)
\label{PGDt}
\EEA
with diffusion coefficients $g^{\Gamma \Gamma}, g^{\Delta \Delta}$, mutation coefficients $m^\Gamma$, $m^\Delta$, and selection coefficients $s_\Gamma, s_\Delta$ that depend on the variables $\Gamma$ and $\Delta$. We obtain the diagonal diffusion coefficients
\begin{eqnarray}
\nonumber g^{\Gamma\Gamma}&=&\sum_{\a,\b}\frac{\partial\Gamma}{\partial x^\a} \ \frac{\partial\Gamma}{\partial x^\b} \ g^{\a\b}\\
\nonumber&=&\sum_{\a,\b}\Tr(\a) \Tr(\b) \left[-x^\a x^\b (1-\delta_\a^\b)+ x^\a(1-x^\a)\delta_\a^\b\right]\\
&=&\ol{(\Tr-\Gamma)^2}=\Delta,
\label{GammaGammaresponse}\\
\nonumber&&\\
\nonumber g^{\Delta\Delta}&=&\sum_{\a,\b}\frac{\partial \Delta}{\partial x^\a}\frac{\partial \Delta}{\partial x^\b}g^{\a\b}\\
\nonumber&=&\sum_{\a,\b}(\Tr(\a)^2-2\ol{\Tr}\Tr(\a))(\Tr_\b^2-2\ol{\Tr}\Tr(\b)) \left[-x^\a x^\b (1-\delta_\a^\b)+ x^\a(1-x^\a)\delta_\a^\b\right]\\
&=& \ol{(\Tr-\Gamma)^4}-\Delta^2
\approx2\Delta^2. 
\label{DeltaDeltaresponse}
\end{eqnarray}
These diffusion coefficients reflect stochastic changes in trait mean and diversity by sampling. It is clear that the fluctuation amplitude (\ref{GammaGammaresponse}) for the trait mean is set by the trait diversity. The corresponding amplitude (\ref{DeltaDeltaresponse}) for the trait diversity is specific to asexual evolution: sampling of a set of complete genotypes with trait values $E_\a$ from a Gaussian distribution ${\cal W} (E)$ with variance $\Delta$ leads to a distribution of sample variances with variance $2 \Delta^2$. This relation changes in recombining populations, where sampling is broken down to individual alleles. For more general trait distributions $\W(E)$, the amplitude $g^{\Delta \Delta}$ given by eq. (\ref{DeltaDeltaresponse}) involves higher moments\cite{deVladar:2011bs,Neher:2011wc}; that is, the closed form (\ref{Ptgen}) of the dynamics for $\Gamma$ and $\Delta$ is a truncation. As shown by our numerical results, this truncation leads to accurate approximations for complex quantitative traits, because their actual trait distribution is approximately Gaussian. As we have anticipated in writing eq.~(\ref{PGDt}), off-diagonal diffusion can be neglected by symmetry,
\EQA
\nonumber g^{\Gamma\Delta}& =&\frac{\partial \Delta}{\partial x^\a}\frac{\partial \Gamma}{\partial x^\b}g^{\a\b}\\
\nonumber&=&(\Tr^2(\a)-2\ol{\Tr}\Tr(\a))\Tr(\b) \left[-x^\a x^\b (1-\delta_\a^\b)+ x^\a(1-x^\a)\delta_\a^\b\right]\\
&=& \ol{(\Tr-\Gamma)^3}\approx0. 
\label{GammaDeltaresponse}
\EEA
This coefficient would lead to additional terms, such as 
$(2N)^{-1}(\partial^2/ \partial \Gamma \partial \Delta) g^{\Gamma \Delta} Q(\Gamma, \Delta, t)$.   The mutation coefficients are 
\begin{eqnarray}
\nonumber m^\Gamma&=&\sum_{i=1}^\ell \frac{\partial\Gamma}{\partial y_i} \mu(1-2y_i)\\
&=&\sum_{i=1}^\ell E_i \mu(1-2y_i)=-2\mu(\Gamma- \Gamma_0 )
\label{MutFieldGamma}\\
\nonumber &&\\
\nonumber m^\Delta &=&\sum_{i=1}^\ell \frac{\partial \Delta}{\partial y_i}\mu (1-2y_i)+\sum_{i\neq j}\frac{\partial \Delta}{\partial y_{ij}}\mu (y_i+y_j-4y_{ij})\\
&=&- 4 \mu (\Delta- E_0^2)-\frac{\Delta}{N}+\mathcal{O}(\theta^2),
\label{MutFieldDelta}
\end{eqnarray}
where $\Gamma_0$ and $E_0$ are given by eq.~(\ref{E0}). The term $\Delta/N$ in (\ref{MutFieldDelta}) appears due to the nonlinear dependence of the trait diversity on the allele frequencies $y_i$ (see, e.g., Chapter 4 of ref.~\cite{Gardiner:2004tx}). Finally, the selection coefficients are the gradient of the phenotypic fitness landscape (\ref{FGammaDelta}), 
\EQ
s_\Gamma = \frac{\partial}{\partial \Gamma} \, F(\Gamma, \Delta),
\hspace{1cm}
s_\Delta = \frac{\partial}{\partial \Delta} \, F(\Gamma, \Delta).
\EE

The two-dimensional diffusion equation (\ref{PGDt}) gives a closed, analytical description of trait evolution under complete genetic linkage. As we show in the next section, it provides numerically accurate results at least for the marginal distributions $Q(\Gamma)$ and $Q(\Delta)$ over a wide range of evolutionary parameters. However, it has the same basic difficulty as the genotypic Kimura equation (\ref{Ptgen}): it does not have an equilibrium solution, because the mutation coefficient field is non-integrable, 
$\partial ((g^{\Gamma \Gamma})^{-1} m^\Gamma)/\partial \Delta - \partial ((g^{\Delta \Delta})^{-1} m^\Delta)/\partial \Gamma \neq 0$. In the Appendix, we show that equation (\ref{PGDt}) leads instead to a non-equilibrium stationary distribution $Q_{\rm stat} (\Gamma, \Delta)$, which is shown in Fig.~2. This distribution satisfies the scaling relation 
\EQ
Q_{\rm stat} (\Gamma, \Delta) = \ell^{-3/2} \, \hat Q_{\rm stat} (\ell^{-1/2} (\Gamma - \langle \Gamma \rangle), \ell^{-1} \Delta) 
\label{Qstat}
\EE
for large values of $\ell$, with ensemble averages $\langle \Gamma \rangle$ and $\langle \Delta \rangle$ of order $\ell$. 
\begin{figure}
\includegraphics[width=0.4\textwidth]{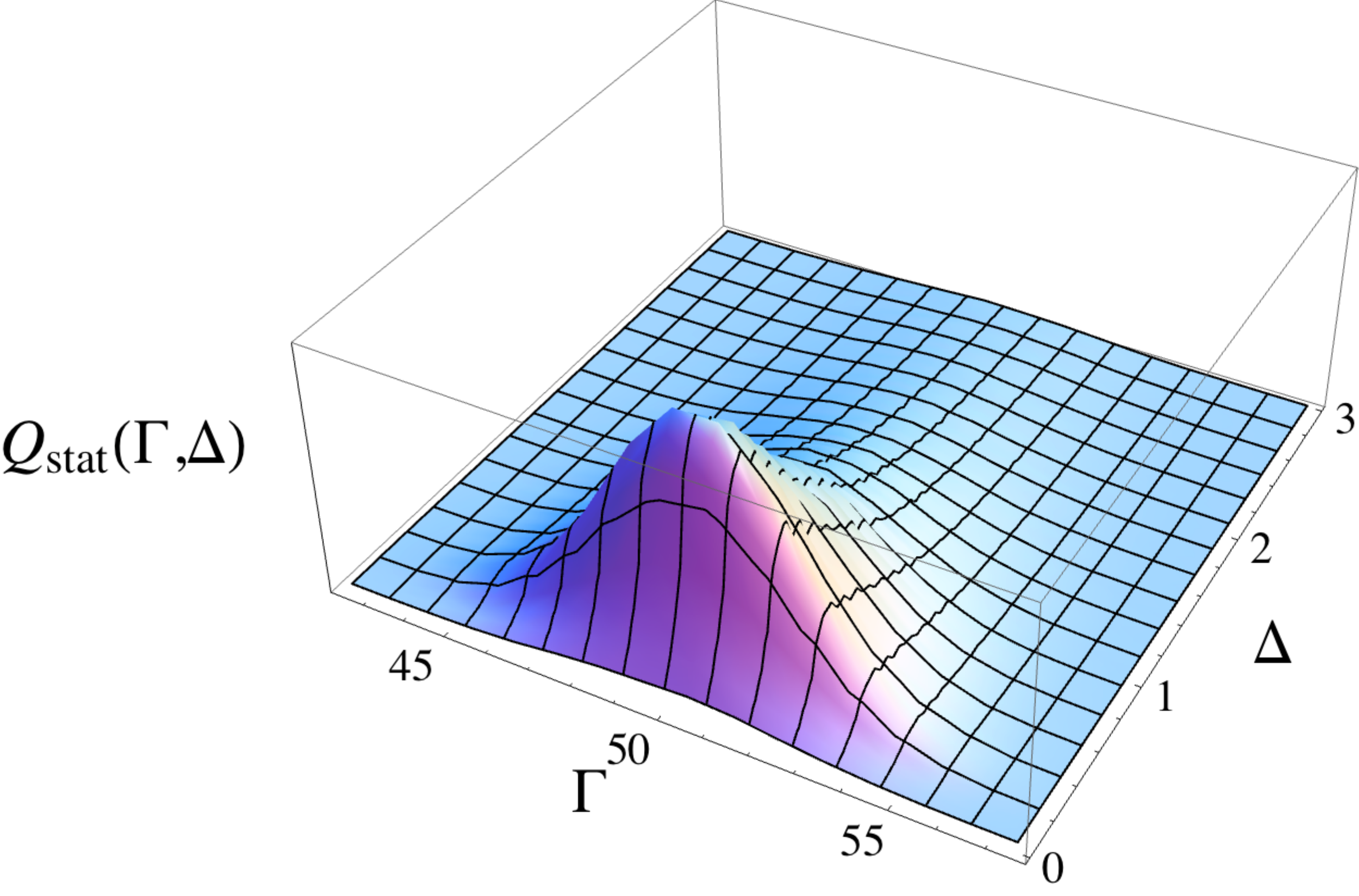}
\caption{{\bf \small Non-equilibrium stationary trait distribution under complete genetic linkage. } \small
Stationary joint distribution of trait mean and diversity, $Q_{\rm stat} (\Gamma, \Delta)$, for a non-recombining population in a quadratic fitness landscape. The figure shows simulation results for a quantitative trait with $\ell =100$ constituent sites of equal effect. The distribution $Q_{\rm stat} (\Gamma, \Delta)$ is Gaussian in the $\Gamma$ direction, but strongly non-Gaussian in the $\Delta$ direction (the resulting marginal distributions are shown in Fig.~3). It maintains a stationary probability current, which is shown in Fig.~7. Other system parameters:  neutral sequence diversity $\theta = \mu N=0.0125$, scaled fitness landscape $2N f(E) = c (E-E^*)^2 /  E_0^2$ of strength $c = 2.5$ with a fitness optimum $E^*=0.5 \Gamma_0$. 
}
\end{figure}
According to this relation, the average $\langle \Gamma \rangle$ and the fluctuations $\hat \Gamma \equiv \Gamma - \langle \Gamma \rangle$ of the trait mean in the stationary ensemble scale in accordance with the central limit theorem,
\EQ
\langle \Gamma \rangle \sim \ell, 
\hspace{1cm}
\langle \hat \Gamma^n \rangle \sim \ell^{n/2} \;\;\, (n = 2,3,\dots),
\label{Gamma_scaling}
\EE
which implies that fluctuations become subleading in the large-$\ell$ limit, $\Gamma = \langle \Gamma \rangle \pm O(\ell^{1/2})$. This scaling also occurs in sexual populations. It is analogous to the thermodynamic limit for macroscopic systems, which is familiar in statistical thermodynamics~\cite{deVladar:2011bs}. However, the average $\langle \Delta \rangle$ and the fluctuations $\hat \Delta \equiv \Delta - \langle \Delta \rangle$ of the trait diversity scale in a different way, 
\EQ
\langle \Delta \rangle \sim \ell,
\hspace{1cm}
\langle \hat \Delta^n \rangle \sim \ell^n \;\;\; (n = 2,3,\dots).
\label{Delta_scaling}
\EE
This scale-invariance of the trait diversity statistics in large-$\ell$ limit is a consequence of coherent, genome-wide linkage disequilibrium fluctuations in the absence of recombination. It is generated  by sampling from a set of genotypes with trait values $E_\a$ from a distribution $\W(E)$ with variance $\Delta \sim \ell$. There is no central-limit theorem, because the number of these genotypes grows only weakly with $\ell$~\cite{shraiman:2012}. In contrast, fast recombination generates a number of genotypes that grows exponentially with $\ell$, 
which leads to the standard scaling $\langle \hat \Delta^n \rangle \sim \ell^{n/2}$ given by the central limit theorem (see \cite{deVladar:2011bs} and the discussion in the next section). These differences in fluctuation statistics are mirrored by the properties of population genealogies: for asexual evolution, there is a single genome-wide genealogy of all genotypes. Standard coalescence theory then predicts diversity fluctuations distributed exponentially, with variance proportional to the square of the coalescence time, which of order $N^2$, and to the square of the genome-wide mutation rate, which in turn is proportional to $\ell^2$. In contrast, recombination generates many parallel genealogies, which average out the diversity fluctuations.

Because the joint evolution of trait mean and diversity is a non-equilibrium process, the diffusion equation (\ref{PGDt}) does not have a simple analytical solution. We now project this dynamics further onto its marginals for $\Gamma$ and $\Delta$. The ensemble distributions $Q(\Gamma,t)$ and $Q(\Delta, t)$ follow coupled one-dimensional diffusion equations, which turn out to have analytical equilibrium solutions.

\subsection{Marginal evolution of the trait mean}
By integrating over the trait diversity in eq.~(\ref{PGDt}), we obtain a one-dimensional diffusion equation for the trait mean. 
This integration amounts to replacing the variable $\Delta$, which appears in the diffusion coefficient $g^{\Gamma \Gamma}$ and in the selection coefficient $s_\Gamma$, by its expectation value $\langle \Delta \rangle$. 
The projected equation reads
\EQ
\frac{\partial}{\partial t} \, Q(\Gamma,t)=\left[\frac{\eff{g}^{\Gamma\Gamma}}{2N}\frac{\partial^2}{\partial\Gamma^2} -\frac{\partial}{\partial\Gamma}\left({m}^\Gamma + \tilde g^{\Gamma \Gamma} \eff{s}_\Gamma \right)\right]Q(\Gamma,t)
\label{Gamma.gen.Fok}
\EE
with the effective diffusion coefficient
\EQ
\tilde g^{\Gamma \Gamma} = \langle \Delta \rangle,
\label{gGammaGamma}
\EE
the mutation coefficient $m^\Gamma = -2\mu(\Gamma- \langle \Gamma \rangle_0)$ given by eq.~(\ref{MutFieldGamma}), and the selection coefficient $\tilde s_\Gamma$, which is the gradient of the effective fitness landscape
\EQ
\tilde F (\Gamma) =\bar f (\Gamma, \langle \Delta \rangle) = f(\Gamma) + \frac{1}{2} \langle \Delta \rangle \, f'' (\Gamma). 
\label{FtildeG}
\EE
This equation has an equilibrium solution 
\EQ
Q_\eq (\Gamma) =  \frac{1}{Z} \,  \tilde Q_0(\Gamma) \exp \!\big [2 N \tilde F(\Gamma) \big], 
\label{QG_eq}
\EE
with
\EQ
\tilde Q_0(\Gamma) \simeq \sqrt{\frac{2 \theta}{ \pi \langle \Delta \rangle}}  \, \exp \left [ - \frac{1}{2} \frac{(\Gamma - \Gamma_0)^2}{ \langle \Delta \rangle / 4\theta } \right ]
\label{QG_0}
\EE
and $\Gamma_0$ given by eq.~(\ref{E0}). The Gaussian form of $\tilde Q_0(\Gamma)$ is valid for sufficiently large values of $\ell$. It implies the scaling form (\ref{Gamma_scaling}) of average and fluctuations of $\Gamma$, in accordance with the central limit theorem. Since $\Gamma$ depends only on the allele frequencies of the constituent loci and not on their linkage correlations, the diffusion equation (\ref{Gamma.gen.Fok}) and the form of  its solution (\ref{QG_eq}), (\ref{QG_0}) are valid regardless of recombination. However, the distribution $\tilde Q_0 (\Gamma)$ depends on the average diversity $\langle \Delta \rangle$ under selection, 
which enters the effective diffusion coefficient (\ref{gGammaGamma}). Hence, $\tilde Q_0 (\Gamma)$ differs from the neutral distribution $Q_0(\Gamma)$. Because $\langle \Delta \rangle$ depends on recombination (see below), the statistics of the trait mean also acquires a small but systematic dependence on the recombination rate.

\subsection{Marginal evolution of the trait diversity}

For non-recombining populations, we obtain a one-dimensional diffusion equation for the trait diversity from eq.~(\ref{PGDt}), 
\EQ
\frac{\partial}{\partial t} \, Q(\Delta,t) =\left[\frac{1}{2N}\frac{\partial^2}{\partial\Delta^2} {g}^{\Delta\Delta}-\frac{\partial}{\partial\Delta}\left({m}^\Delta + g^{\Delta \Delta} \eff{s}_\Delta \right)\right]Q(\Delta,t)
\label{Delta.gen.Fok}
\EE
with the diffusion coefficient $g^{\Delta \Delta} = 2 \Delta^2$ given by ({\ref{DeltaDeltaresponse}), the mutation coefficient $m^\Delta = - 4 \mu (\Delta- E_0^2)-{\Delta}/{N}$  given by (\ref{MutFieldDelta}), and the selection coefficient $\tilde s_\Delta$, which is the gradient of the effective fitness landscape
\EQ
\tilde F (\Delta) = \frac{1}{2} \langle f'' (\Gamma) \rangle \, \Delta.
\label{FDtilde}
\EE
This equation has an equilibrium solution 
\EQ
Q_\eq (\Delta) =  \frac{1}{Z} \,  Q_0(\Delta) \exp \!\big [2N \tilde F(\Delta) \big], 
\label{QD_eq}
\EE
where $Q_0(\Delta)$ is the neutral equilibrium 
\begin{eqnarray}
 Q_0(\Delta)=\frac{1}{Z_0} \Delta^{-3-4\theta} \exp \left [ -\frac{4\theta E_0^2}{\Delta} \right ] 
 \hspace{1cm} \mbox{(no recombination).}
\label{Delta.Dist.Link}
\end{eqnarray}
with the normalization $Z_0 = ( 2\theta E_0^2)^{-2 -4 \theta} \Gamma_{\rm Euler} (2 +4 \theta)$. This distribution has mean and variance 
\EQA
\langle \Delta \rangle_0 & = & 4\theta  E_0^2 \, (1-4\theta)  + \mathcal{O}(\theta^3), 
\nonumber \\
\langle (\Delta - \langle \Delta \rangle_0)^2 \rangle_0 & = & 4\theta E_0^4 \, (1-8\theta)+\mathcal{O}(\theta^3)
\hspace{0.5cm} \mbox{(no recombination).}
\EEA
It is of the form $Q_0 (\Delta) = \ell^{-1} \hat Q_0 (\ell^{-1} \Delta)$ with a scale-invariant shape function $\hat Q_0$, which implies the coherent scaling (\ref{Delta_scaling}) of diversity mean and fluctuations (see also Appendix). 

The trait diversity equilibrium (\ref{QD_eq}, \ref{Delta.Dist.Link}) can be compared with its counterpart for free recombination. The  equilibrium distribution $Q_\eq (\Delta)$ for the free-recombining traits is also of the form (\ref{QD_eq}), with the neutral distribution $Q_0 (\Delta)$ obtained by projection from the allele frequency distribution (\ref{Py_0}),
\EQA
Q_0 (\Delta)  &=&  \int \updelta \big (\Delta - \sum_{i=1}^\ell E_i^2 y_i (1- y_i) \big ) P_0(y) dy_1 \dots dy_\ell
\nonumber \\
& & \hspace{6cm}
\mbox{(free recombination).}
\label{QD_0free}
\EEA
For sufficiently large $\ell$, this distribution is again Gaussian with mean and variance~\cite{deVladar:2011bs}, 
\EQA 
\langle \Delta \rangle_0 & = &4 \theta  E_0^2 \, (1-4\theta)+\mathcal{O}(\theta^3), 
\nonumber \\
\langle (\Delta - \langle \Delta \rangle_0)^2 \rangle_0 & = &  \theta \sum_{i=1}^\ell E_i^4 \, \left (\frac{1}{6}  - \frac{14 \theta}{9} \right )+\mathcal{O}(\theta^3) \hspace{0.5cm}
\mbox{(free recombination),}
\EEA
which implies the standard scaling of diversity average and fluctuations, $\langle \Delta \rangle \sim \ell$ and $\langle \hat \Delta^n \rangle \sim \ell^{n/2}$ for ($n = 2,3, \dots$). 

In a generic fitness landscape, the equilibrium distributions (\ref{QG_eq}) and (\ref{QD_eq}) for trait mean and diversity depend on each other, and consistent joint solution has to be obtained iteratively. Mean and diversity decouple in a linear fitness landscape~\cite{Barton:2009genetics}, and the dynamics of the diversity is still autonomous in a quadratic fitness landscape. This case will be discussed in the next section. 

We test our analytical results by simulations of a Fisher-Wright process under stabilizing selection and at neutrality. We evolve a population of $N$ individuals with genomes $\a^{(1)}, \dots, \a^{(N)}$, which are bi-allelic sequences of length $\ell$. A genotype $\a$ defines a phenotype $E(\a) = \sum_{i=1}^\ell E_i a_i$; the phenotypic effects $E_i$ drawn from various distributions. In each generation, the sequences undergo point mutations with a rate $\tau \mu$ per generation (where $\tau$ is the generation time). The sequences of next generation are then obtained by multinomial sampling; the sampling probability is proportional to $[1 + \tau f(E(\a)]$ with the fitness $f(E) = -c_0 \, (E - E^*)^2$ (details are given in the next section). For sexual populations, we permute the alleles $a_{i,1}, \dots, a_{i,N}$ at each genomic site $i$ between the individuals in each generation, which amounts to recombination with an infinite rate. As shown in Fig.~3, the analytical equilibrium distributions $Q_\eq (\Gamma)$ and $Q_0(\Gamma)$ given by eqs.~(\ref{QG_eq}, \ref{QG_0}), as well as  the distributions $Q_\eq(\Delta)$ and $Q_0 (\Delta)$ given by eqs.~(\ref{Delta.Dist.Link}, \ref{QD_0free}), are in good agreement with simulation results. Stabilizing selection shifts the average and reduces the variance of the distribution $Q(\Gamma)$, and it reduces average and variance of the distribution $Q(\Delta)$ compared to neutral evolution. The dependence of these effects on the strength of selection is analyzed in the next section. 

\begin{figure}
\includegraphics[width=\textwidth]{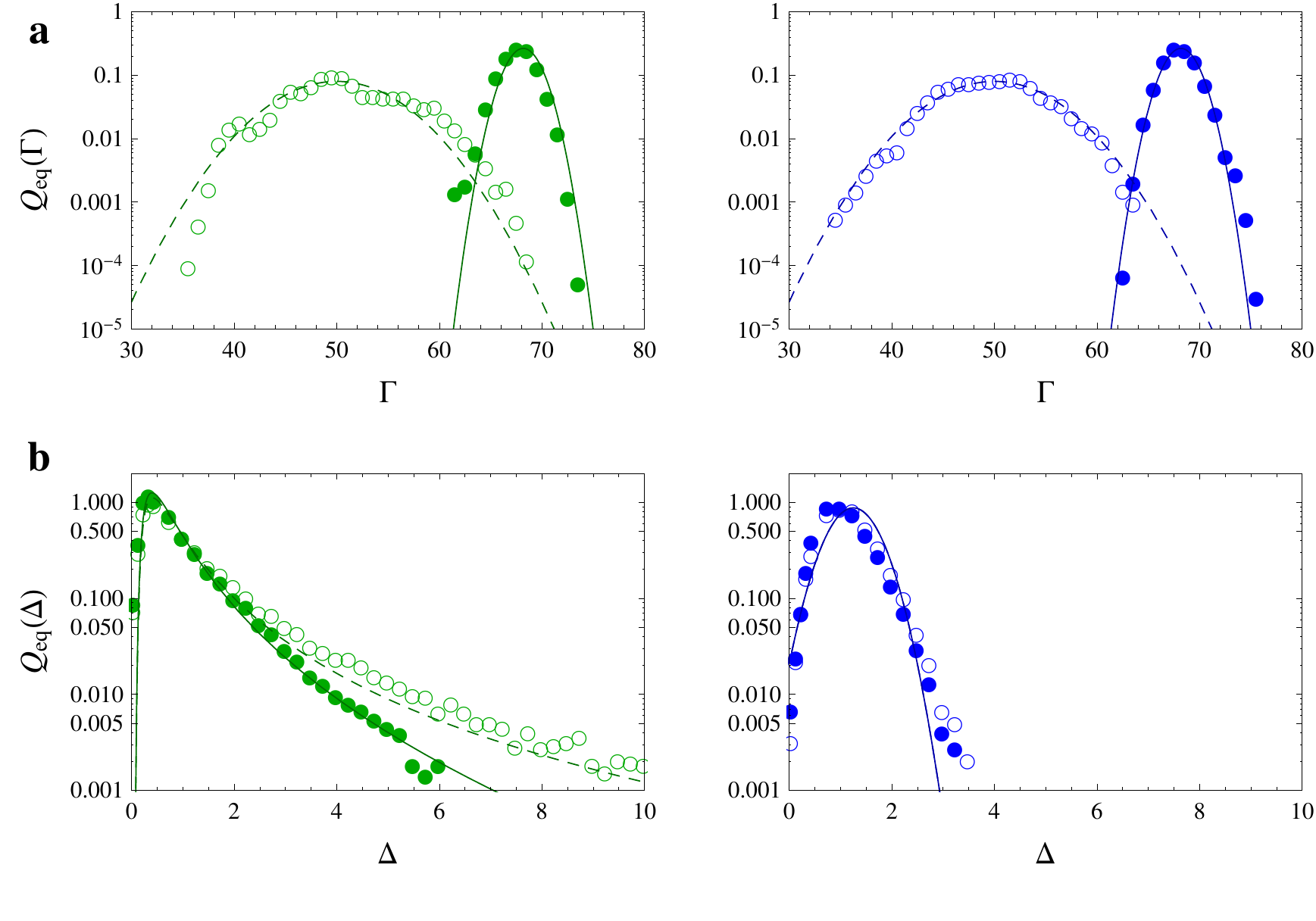}
\caption{\label{fig:traitdist}
{\bf \small Equilibrium trait distributions under stabilizing selection and at neutrality.} \small
(a)~Equilibrium distribution $Q_\eq(\Gamma)$ of the trait mean in a quadratic fitness landscape (filled circles) and corresponding neutral equilibrium $Q_0 (\Gamma)$ (empty circles) (green: no recombination, blue: free recombination).
(b)~Equilibrium distribution $Q_\eq(\Delta)$of the trait diversity (green: no recombination, blue: free recombination).
Theory predictions for these distributions are shown as solid and dashed lines. System parameters: $\ell = 100$ trait loci of equal effect $E_i=1\ (i=1\dots\ell)$, neutral sequence diversity $\theta = \mu N=0.0125$, scaled fitness landscape $2N f(E) = -c (E-E^*)^2 / E_0^2$ of strength $c = 5$ with a fitness optimum $E^*=0.7 L$. Results for other effect distributions are shown in Fig.~6.}
\end{figure}

\section{Trait equilibria under stabilizing selection}
\label{sec_sel}

We now apply our statistical model to quantitative traits under stabilizing selection, a scenario described by evolutionary equilibrium in a quadratic fitness landscape,
\EQ
f(\Tr) = f^* - {c_0} \, (\Tr - \Tr^*)^2  , \qquad (c_0>0). 
\label{eq_quadFit}
\EE
This scenario is probably a reasonable approximation for many actual traits, which have high-fitness values in a certain range around their optimum value $E^*$~\cite{deVladar:2011bs,Barton:1986,Barton:2009genetics}. For example, it applies to the expression level of a gene: small changes in expression may be buffered by compensatory changes in the regulatory network and will affect fitness only weakly, but larger changes are often deleterious, as it is evident from the large number of genetic disorders associated with gene copy number variation. 

Stabilizing selection changes the distribution of trait values in a population, $\w(E)$, which can be parametrized by changes in the trait mean $\Gamma$ and the diversity $\Delta$. Statistical theory describes the expectation values of these changes in an ensemble of populations. At a qualitative level, the main effects are already clear from the previous section: stabilizing selection decreases the average squared distance from the fitness optimum, $\langle \Lambda \rangle^2  \equiv (\langle \Gamma \rangle - E^*)^2$, the average equilibrium divergence between populations, $\langle (\Gamma_1 - \Gamma_2)^2 \rangle = 2 \langle \hat \Gamma^2 \rangle$, and the average diversity, $\langle \Delta \rangle$. We now derive analytical expressions for these effects in non-recombining populations and under free recombination, and we analyze their dependence on the selection strength $c_0$ (the fitness maximum $f^*$ is an arbitrary constant, because the evolution equation (\ref{PGDt}) depends only on fitness gradients); see also refs.~\cite{Blumer:1972,Barton:1986,Wright:1937,Wright:1935Jg,deVladar:2011bs} for  effect of stabilizing selection on free-recombining macroscopic traits. Compared to a generic fitness landscape, the analysis is somewhat simplified for a quadratic fitness landscape (\ref{eq_quadFit}), because the mean population fitness separates,
\EQ
F(\Gamma, \Delta) =f^\star- c_0  (\Gamma - E^*)^2 - {c_0}\Delta.
\label{FGammaDeltastabsel}
\EE

For a quantitative analysis, it is useful to measure phenotypes in a natural unit, which avoids the arbitrariness of fixed units (such as centimeters or inches for body height).  Here we express trait values in units based on the effect amplitude (\ref{E0}),
\EQ
e \equiv \frac{E}{E_0}, 
\hspace{1cm}
\gamma \equiv \frac{\Gamma}{E_0}, 
\hspace{1cm}
\delta \equiv \frac{\Delta}{E_0^2},
\label{scaling}
\EE
and in the same way $e^* \equiv E^* / E_0$, $\lambda \equiv \Lambda / E_0$ and $\hat \gamma \equiv \hat \Gamma / E_0$. These scaled values are pure numbers (we distinguish them by use of lower case letters from the raw data). 
The scaling (\ref{scaling}) has a straightforward biological interpretation: $E_0^2$ is the trait variance in an ensemble of random genotypes, which would result from neutral evolution in the weak-mutation regime,
\EQ
E_0^2  =  \lim_{\mu \to 0} \, \langle (\Gamma - \langle \Gamma \rangle)^2 \rangle_0 .
\EE
We also use the effect amplitude to define the scaled strength of stabilizing selection,
\EQ
c \equiv 2 N E_0^2 \, c_0,
\label{cscaling}
\EE
which can be interpreted as  the difference between the fitness maximum $f^*$ and the average fitness in the random ensemble, 
\EQ
 c = 2N f^* - \lim_{\mu \to 0} \, \langle 2N \bar f \rangle_0 \; ,
\EE
where fitness (growth rate) is measured per $2N$ generations and we have assumed that selection does not shift the trait average (i.e., $\Tr^\star=\langle \Gamma \rangle_0 $). Such fitness differences are referred to as {\em genetic load}, which is discussed in section \ref{load}.

\begin{figure}
\includegraphics[width=\textwidth]{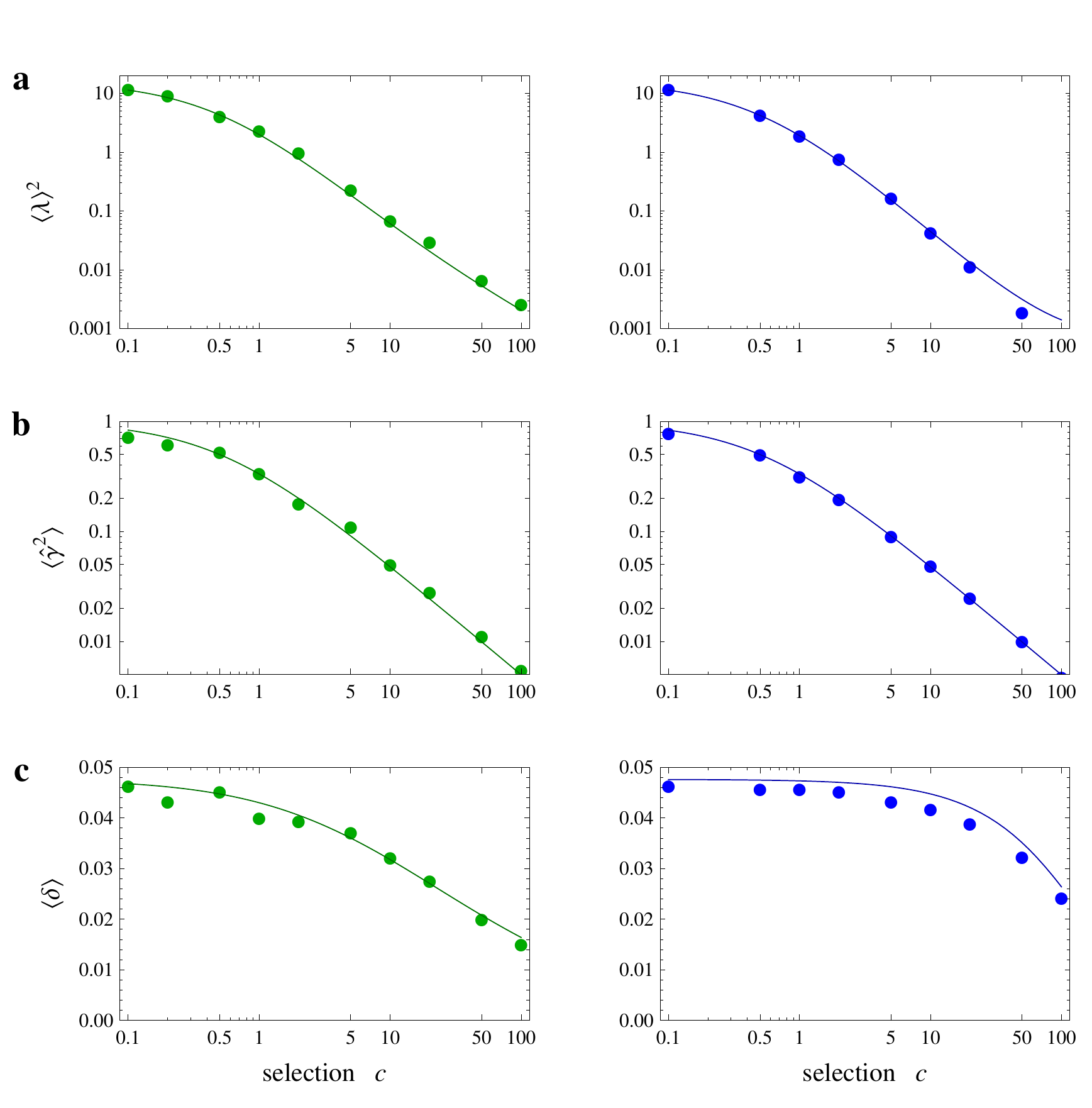}
\caption{\label{fig:traitmoments-selection}
{\bf \small Trait moments under stabilizing selection.} \small
(a)~The squared average distance of the trait mean from the fitness optimum, $\langle \lambda\rangle^2$,
(b)~the variance of the trait mean, $\langle \hat \gamma^2 \rangle$, which equals half the average equilibrium divergence, and
(c)~the average diversity $\langle\delta\rangle$
are plotted against the selection strength $c$ (green: no recombination, blue: free recombination). Other system parameters are as in Fig.~3. All quantities are scaled by the effect amplitude $E_0$. In both recombination regimes, the effect of stabilizing selection on the trait diversity is seen to be smaller than on the trait mean. }
\end{figure}

\subsection{Trait average under stabilizing selection}

In a quadratic fitness landscape, the equilibrium distribution of the trait mean  is Gaussian for sufficiently large values of $\ell$,
\EQ
Q_\eq (\Gamma) = \frac{1}{Z_\Gamma} \tilde Q_0 (\Gamma) \exp [2N f (\Gamma)] =\frac{1}{Z_\Gamma} 
\exp \left [ \frac{2 \theta}{\langle \delta \rangle} (\gamma - \gamma_0)^2 - c (\gamma - e^*)^2 \right ]
\EE
with $\gamma_0 = \sum_{i=1}^\ell e_i/2$, as given by eqs.~(\ref{FtildeG}), (\ref{QG_eq}), and (\ref{QG_0}) and $Z_\Gamma$ as the appropriate normalization factor. This distribution has the scaled moments 
\EQA
\langle \lambda \rangle^2 \equiv \big( \langle \gamma \rangle - e^*\big)^2 & = & 
\langle \lambda \rangle_0^2 \; \frac{1}{(1+c\Mdel/2\theta)^2} \; ,
\nonumber \\
\langle \hat \gamma^2 \rangle \equiv \langle \gamma^2 \rangle - \langle \gamma \rangle^2  & = &  
\langle \hat \gamma^2 \rangle_0 \, \frac{\langle \delta \rangle}{\langle \delta \rangle_0}  \; \frac{1}{(1+c\Mdel/2\theta)} \;. 
\label{moments_mean}
\EEA
with $\langle \lambda \rangle_0 = \gamma_0 - e^*$ and $\langle \hat \gamma^2 \rangle_0 =1 - 4 \theta + \mathcal{O} (\theta^2)$. In the regime of weak selection ($c \ll 1$), these moments depend on the selection strength $c$ in a universal way, 
\EQA
\langle \lambda \rangle^2 & = & 
\langle \lambda \rangle_0^2  \, (1 - 4 c)  +\mathcal{O}(\DLc^2,c/\ell,\DLc\theta),
\nonumber \\
\langle \hat \gamma^2 \rangle &  = & 
\langle \hat \gamma^2 \rangle_0  \, (1 - 2 c) +\mathcal{O}(\DLc^2,c/\ell,\DLc\theta),
\EEA
because the effect of selection on the trait diversity is subleading ($\langle \delta \rangle / \langle \delta \rangle_0 = 1 + \mathcal{O}(c/\ell,\DLc\theta)$, see eq.~(\ref{Dpert-free}) below). For larger values of $c$, these moments acquire a noticeable dependence on $\langle \delta \rangle$, and thereby on the recombination rate. For asexual populations, we obtain the strong-selection regime ($c \theta \gg 1$) 
\EQA
\langle \lambda \rangle^2 & = & 
\langle \lambda \rangle_0^2 \, \frac{\theta}{c} \; [1 + \mathcal O({\theta^{1/2}}{c^{-1/2}})], 
\nonumber \\
\langle \hat \gamma^2 \rangle &  = & 
\frac{1}{2c} \; [1 + \mathcal O(\theta, c^{-1/2})]
\hspace{1.5cm} \mbox{(no recombination)},
\EEA
where we have used  eq.~(\ref{delta_no}) below. Evaluating this regime does not make sense in the free-recombination approximation, because if epistatic selection is strong, the assumption of linkage equilibrium breaks down for any finite recombination rate.

\subsection{Trait diversity under stabilizing selection}

The equilibrium distribution of the trait diversity is 
\EQ
Q_\eq (\Delta) =\frac{1}{Z_\Delta}  Q_0(\Delta) \exp(-2N c_0 \Delta) 
\EE
with $Q_0(\Delta)$ given by (\ref{Delta.Dist.Link}) and (\ref{QD_0free}) and $Z_\Delta$ as the appropriate normalization constant. This distribution does not depend on the statistics of the trait mean and determines the scaled average diversity in asexual populations by, 
\EQA
\langle \delta \rangle & = &
\frac{1}{Z_\delta} \int \delta^{-2-4\theta} \exp \left ( -\frac{4\theta}{\delta} - c \delta \right ) \, d \delta  
\nonumber \\
&  = &  \sqrt{\frac{4\theta}{ c}} \, \frac{k_{1+4 \theta}[4\sqrt{\theta  \DLc}]}{k_{2+4 \theta} [4\sqrt{\theta \DLc}] }
\nonumber \\
& \equiv  & \langle \delta \rangle_0 \, \big [ 1 + {\cal G} (\theta c) \big ]
\hspace{3cm} \mbox{(no recombination),}
\label{delta_no}
\EEA
where $k_n (z)$ denotes the modified Bessel function of the second kind. The average diversity of free-recombining traits reads (see also ref.~\cite{deVladar:2011bs}),
\EQA
\langle \delta \rangle & = &
\sum_{i=1}^\ell \frac{e_i^2}{Z_{\delta_i}} \int_0^{e_i^2/4} \frac{ (\delta_i/e_i^2)^{2\theta}}{\sqrt{1 -4 (\delta_i / e_i^2)}} \, \exp ( - c \delta_i ) \, d \delta_i 
\nonumber \\
& = & \frac{\theta}{2}\sum_{i=1}^\ell   e_i^2 \ \frac{ F_1[1+2\theta,3/2+2\theta,-c\ e_i^2/4]}{  F_1[2\theta,1/2+2\theta,-c \ e_i^2/4]}
\nonumber \\
& = & 
\frac{\theta}{2}\int d\DLtr \  \DLK(\DLtr) \DLtr^2\  \frac{ F_1[1+2\theta,3/2+2\theta,-(c/\ell) \ \DLtr^2/4]}{  F_1[2\theta,1/2+2\theta,-(c/\ell) \ \DLtr^2/4]} 
\nonumber \\
& \equiv & \langle \delta \rangle_0 \, \left [ 1+ {\cal G}_{\rm free} \!\left ( \frac{c}{\ell}, \kappa \right ) \right ]
\hspace{2.5cm} \mbox{(free recombination),}
\label{delta_free}
\EEA
where $F_1[a,b,z]$ is the regularized confluent hypergeometric function and we have introduced the effect density 
\EQ
\kappa (\epsilon) \equiv \frac{1}{\ell} \sum_{i=1}^\ell \updelta( \epsilon - \ell^{1/2} e_i). 
\label{epsilon}
\EE
We can again expand these expressions to leading order in $c$,
\EQA
\langle \delta \rangle  & =  & \langle \delta \rangle_0 \, 
\big [ 1 - 4 \theta c + \mathcal{O} (\theta^2 c^2) \big ]
\hspace{2cm}  \mbox{(no recombination),}
\\
\label{Dpert-free}\langle \delta \rangle  & =  & \langle \delta \rangle_0 \, 
\left [ 1 - \frac{2 \kappa_4}{3 \kappa_2^2} \frac{c}{\ell} + \mathcal{O} \left ( \frac{c^2}{\ell^2}, \frac{c \theta}{\ell} \right ) \right ]
\hspace{0.5cm}  \mbox{(free recombination),}
\EEA
where $\kappa_n$ denotes the $n$-th moment of the distribution $\kappa (\epsilon)$ ($n = 1,2, \dots$). 
For asexual populations, we obtain the strong-selection regime ($c \theta \gg 1$) 
\EQ
\langle \delta \rangle   =   \langle \delta \rangle_0 \, 
\left [\frac{1}{(4\theta c)^{1/2}} + \mathcal{O} \left (\frac{1}{\theta c} \right ) \right ]
\hspace{1.5cm}  \mbox{(no recombination)}.
\EE
Again, evaluating this regime does not make sense in the free-recombination approximation, because approximate linkage equilibrium cannot be maintained at any finite recombination rate. For $c/\ell \gg 1$, selection changes even qualitatively: it becomes balancing at individual trait loci and would act to increase the trait diversity. 

Our analytical results (\ref{moments_mean}), (\ref{delta_no}), and (\ref{delta_free}) for trait equilibria under stabilizing selection are shown in Fig.~4 together with numerical simulations. As expected, the behavior of the trait diversity depends more strongly on the recombination rate than that of the trait mean. However, there is an important and universal feature: stabilizing selection affects the trait diversity always less than its mean. This feature, which will be the basis for a test of stabilizing selection on quantitative traits, is explicitly demonstrated by our solution. As shown by eqs.~(\ref{delta_no}) and (\ref{delta_free}), selection on trait diversity has an effective strength 
\EQA 
\theta c & \ll c \hspace{2.5cm} \mbox{(no recombination),}
\\
c / \ell & \ll c \hspace{2.5cm} \mbox{(free recombination),}
\EEA
which involves a small prefactor compared to the selection strength $c$ acting on divergence. These prefactors reflect different mechanisms of stabilizing selection acting on trait diversity. In asexual populations, selection acts on a distribution of genotypes, which generates a neutral trait diversity by a factor $\theta$ smaller than the neutral trait divergence. In sexual populations, selection acts on individual trait loci, and the mean square trait amplitude of an individual locus by a factor of order $(1/\ell)$ smaller than the mean square amplitude $E_0^2$ of the entire trait.

\section{Fitness and entropy under stabilizing selection}
\label{fitness_entropy}

The distributions of trait mean and diversity derived in the previous section also determine the fitness and entropy statistics in the equilibrium population ensemble. This statistics provides a few biologically relevant numbers: it quantifies how well adapted typical populations are under stabilizing selection, how much adaptation has occurred between neutrality and the adapted state, and how much measurements in one population can predict about another population evolving in the same fitness landscape.

\subsection{Genetic load}
\label{load}

How far away is a population from the fitness peak? This question is answered by the {\em genetic load} 
\EQ
L \equiv f^* - \bar f, 
\EE
which is defined as the difference between the fitness maximum and the mean population fitness (and is conveniently measured in units of $1/2N$)~\cite{Muller:1950,Haldane:1957,Crow:1958,Crow:1965}. In the quadratic fitness landscape (\ref{eq_quadFit}), we can decompose $L$ into a component associated with the trait mean, $2NL_\Gamma \equiv c (\gamma - e^*)^2$, which is generated mainly by substitutions away from the fitness optimum, and the diversity load, $2N L_\Delta \equiv c \delta$, which is generated by trait polymorphisms. Our statistical theory predicts the ensemble average of the genetic load at equilibrium, 
\EQ
\langle 2N L \rangle = c \big ( \langle \lambda \rangle^2 + \langle \hat \gamma^2 \rangle + \langle \delta \rangle \big ) ,
\label{eq.Load}
\EE 
in terms of the leading moments of trait mean and diversity, which are given by eqs.~(\ref{moments_mean}), (\ref{delta_no}), and (\ref{delta_free}). Fig.~5(a) shows that the total load and its two components depend on the strength of selection in a non-monotonic way. For weak selection, the main load component is $\langle 2NL_\Gamma \rangle$,  but $\langle2N L_\Delta \rangle$ dominates for strong selection. This reflects our result that stabilizing selection affects the trait diversity less than its mean.

\begin{figure}[t!]
\includegraphics[width=\textwidth]{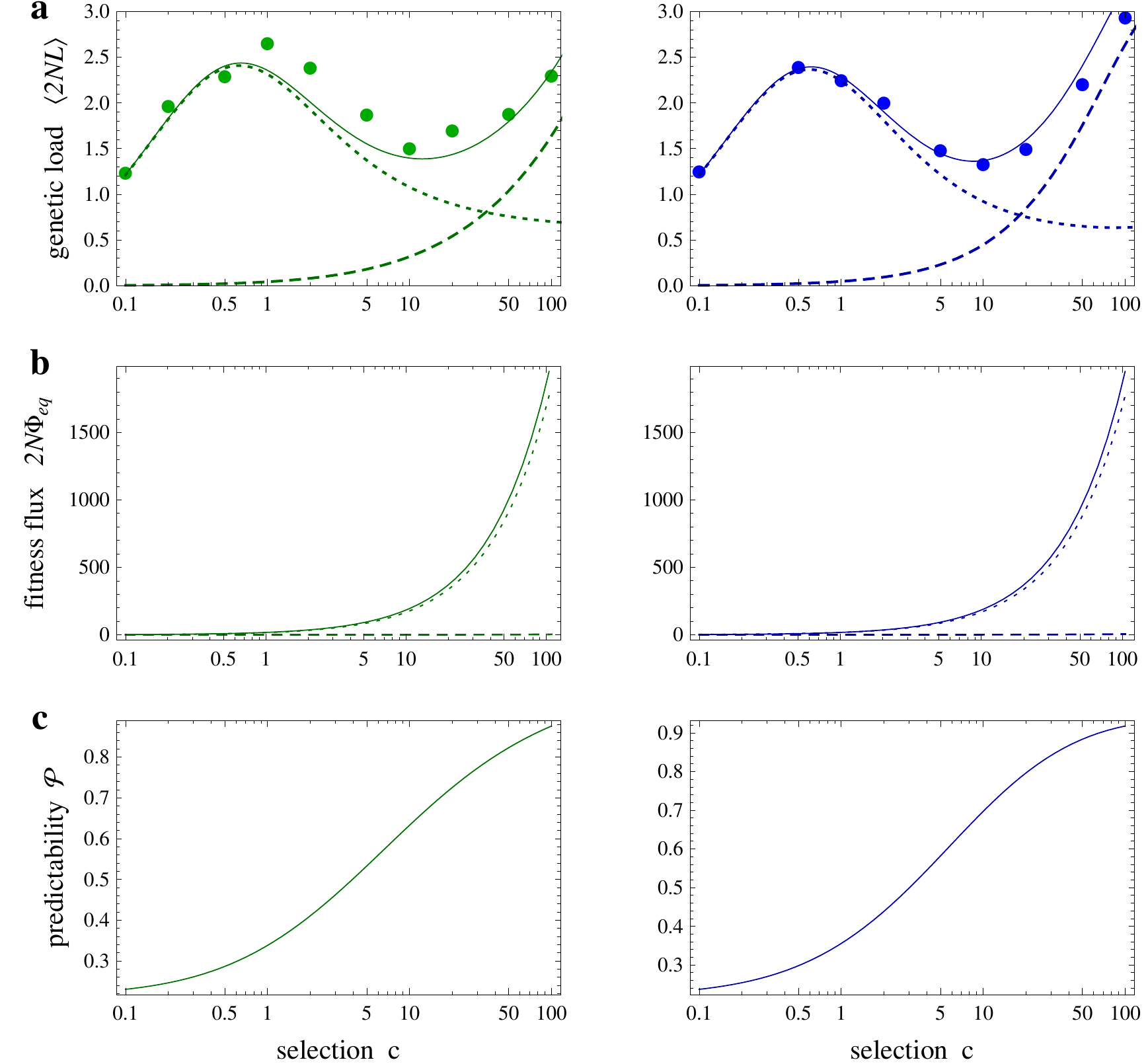}
\caption{
\label{fig:Load}
{\bf \small Genetic load, fitness flux, and predictability of evolution.} \small 
(a)~The average genetic load $\langle L \rangle $
(full lines) with its components $\langle L_\Gamma \rangle $ (dotted lines) and $\langle L_\Delta \rangle $ (dashed lines),
(b)~the equilibrium fitness flux $\Phi_\eq$  with its components $\Phi_{\eq, \Gamma}$ (dotted lines) and $\Phi_{\eq, \Delta}$ (dashed lines), and 
(c)~the predictability ${\cal P}$ 
are plotted  against the selection strength $c$ (green: no recombination, blue: free recombination; fitness is measured in units of $1/2N$). See definitions in eqs.~(\ref{eq.Load}), (\ref{Phi}), and (\ref{calP}). Other system parameters are as in Fig.~3.}
\end{figure}

\subsection{Free fitness and fitness flux}
\label{flux}

How far away is a population ensemble from neutral evolution? This can be measured in two ways: by the difference in average scaled fitness between that ensemble and the neutral ensemble 
\EQ
\langle 2N \bar f \rangle_Q - \langle 2N \bar f \rangle_0 = \langle 2N L \rangle_0- \langle 2N L \rangle_Q, 
\EE
and by the relative entropy or Kullback-Leibler distance between the ensemble under selection and the neutral ensemble,
\EQ
H(Q|Q_0) \equiv \int d\Gamma \, d \Delta \, Q(\Gamma, \Delta) \log \left [ \frac{Q(\Gamma, \Delta)}{Q_0(\Gamma, \Delta)} \right ].
\EE
The difference between scaled fitness and relative entropy is called {\em free fitness}, 
\EQ 
{\cal F} (Q) \equiv \langle 2N \bar f \rangle_Q - H(Q|Q_0);
\label{freeF} 
\EE
see refs.~\cite{Iwasa:1988ws, Berg:2004dz,Sella:2005da,Barton:2009di,Mustonen:2010ig}. 
This quantity is of particular importance, because it satisfies a growth principle similar to Boltzmann's $H$-theorem in statistical physics: for any evolutionary process in a time-independent fitness landscape which has an equilibrium, the free fitness ${\cal F} (Q(t))$ increases monotonically with time and has its maximum at equilibrium~\cite{Iwasa:1988ws,Sella:2005da,Mustonen:2010ig}. Here we approximate the stationary trait distribution under stabilizing selection by the product of its equilibrium marginal distributions, $Q_{\rm stat} (\Gamma, \Delta) \approx Q_\eq (\Gamma) Q_\eq (\Delta) \equiv Q_\eq$; the same approximation is used for the neutral distribution $Q_0(\Gamma, \Delta)$  (the results in the Appendix show that this is numerically justified). We then obtain the relative entropy 
\EQ
H(Q_\eq | Q_0)  =  - c \big ( \langle \lambda \rangle^2 + \langle \hat \gamma^2 \rangle \big ) - \log Z_\Gamma - c \langle \delta \rangle - \log Z_\Delta
\EE
and the difference in free fitness or {\em fitness flux}
\EQA
2N \Phi_\eq  & \equiv & {\cal F} (Q_\eq) - {\cal F} (Q_0) 
\nonumber \\
& = & - c \big ( \langle \lambda \rangle^2 + \langle \hat \gamma^2 \rangle + \langle \delta \rangle \big ) - H(Q_\eq | Q_0) + c \big ( \langle \lambda \rangle_0^2 + \langle \hat \gamma^2 \rangle_0 + \langle \delta \rangle _0 \big )
\nonumber \\
& = & c \big ( \langle \lambda \rangle_0^2 + \langle \hat \gamma^2 \rangle_0 \big ) + \log Z_\Gamma + c \langle \delta \rangle_0 + \log Z_\Delta, 
\label{Phi}
\EEA
with $\log Z_\Gamma \simeq \langle \lambda\rangle_0^2(\theta-(\theta c)^{1/2})-(1/2) \log c$ and $\log Z_\Delta \simeq -4(\theta c)^{1/2} +(3/4) \log c$. 
The scaled fitness flux $2 N \Phi_\eq$ measures the total amount of adaptation between the neutral equilibrium and the equilibrium under stabilizing selection\footnote{
Fitness flux plays a central role as a measure of adaptation also in non-equilibrium processes, where it is no longer related to free energy changes~\cite{Mustonen:2010ig}.}~\cite{Mustonen:2010ig}. As shown in Fig.~\ref{fig:Load}(b), this flux is always positive and increases with the selection strength $c$. Similarly to the genetic load, it can be decomposed into contributions of the trait mean and the trait diversity, $2N \Phi = 2N \Phi_{\eq,\Gamma} + 2N \Phi_{\eq, \Delta}$. The term $2N \Phi_{\eq,\Gamma} = c \big (\langle \lambda \rangle_0^2 + \langle \hat \gamma^2 \rangle_0 )+\log Z_\Gamma$ is the dominant contribution, again because stabilizing selection affects the trait diversity less than its mean.

\subsection{Predictability of evolution}
\label{predictability}

How informative are trait measurements in one population about the distribution of trait values in a replicate population evolving in the same fitness landscape? To answer this question, we compare the ensemble-averaged Shannon entropy of the phenotype distribution within a population, 
\EQ
\langle S \rangle_\W \equiv \int_{\W}  S(\W) \, Q(\W) 
\label{Save}
\EE
and the Shannon entropy of the ``mixed'' distribution 
\EQ
S(\langle \W \rangle ) \equiv S \big ( \int_{\W}  \W \, Q(\W) \big ),
\label{Smix}
\EE
which is obtained by compounding the trait values of all populations into a single distribution. We 
define the phenotypic {\em predictability} 
\EQ
\P \equiv \exp \big [\langle S \rangle_\W - S( \langle W \rangle) \big ]
\label{calP}
\EE
with $S (\W)  \equiv - \int \W(E) \, \log \W(E) dE$. This quantity measures how much of the total trait value repertoire of all populations is already contained in the trait distribution $\W(E)$ of a single distribution. It is closely related to the expected overlap between the distributions $\W_1 (E)$ and $\W_2 (E)$ of two replicate populations. 

To compute the predictability under stabilizing selection, we approximate the ensemble average in (\ref{Save}) and (\ref{Smix}) by an average over $\Gamma$, using the approximate parametrization $\W(E | \Gamma) \sim  \exp[-(E - \Gamma)^2 / 2 \langle \Delta \rangle]$. We obtain 
\EQ
\P \simeq \left(\frac{\langle \Delta \rangle}{\langle \hat \Gamma^2 \rangle + \langle \Delta \rangle}\right)^{1/2}= \left(\frac{1}{1+\Omega / 4 \theta}\right)^{1/2} 
\label{calP_eq}
\EE
with the dimensionless ratio
\EQ
\Omega \equiv
\frac{ \langle \hat \gamma^2 \rangle /  \langle \hat \gamma^2 \rangle_0 }{ \langle \delta \rangle / \langle \delta \rangle_0 } = 
\left \{
\begin{array}{ll} 
\left [ 1 + \displaystyle{2{c}} \, \big ( 1 + {\cal G} (\theta c) \big ) \right ]^{-1} & \mbox{(no recombination)}, 
\\ 
\left [ 1 + \displaystyle{2{c}} \, \big ( 1 + {\cal G_{\rm free}} (c/\ell, \kappa) \big ) \right ]^{-1} & \mbox{(free recombination)}
\end{array} \right. 
\label{eq_Omega}
\EE
given by eqs.~(\ref{moments_mean}), (\ref{delta_no}), and (\ref{delta_free}). The dependence of $\P$ on the strength of stabilizing selection is shown in Fig.~\ref{fig:Load}{(}c). While the neutral predictability $\P_0 = 4 \theta / (1 + 4 \theta)$ is small, stabilizing selection can generate predictability values $\P$ of order 1. The reason is again because the trait mean is more constrained than the trait diversity. This feature is illustrated in Fig.~1: under selection, a single-population distribution $\W(E)$ fills a larger fraction of the trait range spanned by the cross-population distribution $Q(\Gamma)$ than at neutrality.

It is instructive to compare the phenotypic predictability (\ref{calP}) with the analogous measure for genotypes,
\EQA
\P_g & \equiv & \exp \big [ \langle S \rangle_x - S ( \langle x \rangle )\big ]
\nonumber \\
& = & \exp \left [ \mbox{$\sum_\a$} \big ( \langle x_\a \log x_\a \rangle - \langle x_\a \rangle \log \langle x_\a \rangle \big ) \right ]. 
\label{genpred}
\EEA
For complex traits (i.e., for large values of $\ell$), we find the genotypic predictability
\EQ
\P_g  \simeq  \exp \left [ - \ell \, [\varsigma (c ) - \mathcal{O} (\theta,\ell^{-1}) ]\right ]. 
\label{chargenpred}
\EE
The leading entropy density $\varsigma (c )$ is given by 
\EQ
\varsigma ( c) \simeq \left \{
\begin{array}{ll} 
\log 2 & \mbox{ for $ c \ll 1$,}
\\
\alpha \, E^*/\ell - \int d \epsilon \, \kappa (\epsilon) \, \log (1 + {\rm e}^{\alpha \epsilon}) & \mbox{ for $ c \gg 1$,}
\end{array} \right. 
\label{varsigma}
\EE
where $\kappa (\epsilon)$ is the single-locus effect distribution defined in eq.~(\ref{epsilon}). The constant $\alpha$ is implicitly determined by the condition
\EQ
\int d \epsilon \, \kappa (\epsilon) \, \frac{\epsilon \, {\rm e}^{\alpha \epsilon}}{1 + {\rm e}^{\alpha \epsilon}}  = \frac{E^*}{\ell}.
\EE
To derive this result, we note that $\varsigma ( c)$ is determined by the entropy of the ``mixed'' distribution, $S(\langle x \rangle)$, which can be evaluated in the low-mutation limit $\theta \to 0$. Hence, $\varsigma (c )$ is also independent of recombination, which affects the overlap statistics between genotypes within a population~\cite{shraiman:2012}  and only enters the $\theta$-dependent corrections. Asymptotically for $\theta \ll 1$ and $c \ll 1$, the mixed entropy reduces to the logarithm of the number of sequence states at the constitutive sites, $S(\langle x \rangle) \simeq \ell \, \log 2$. In the strong-selection regime, we can compute this entropy using the canonical formalism of statistical mechanics. We evaluate the partition function under linear selection on the trait, 
 \EQ
Z_g = \prod_{i=1}^\ell \sum_{\sigma_i = 0,1} {\rm e}^{\alpha E_i \sigma_i} 
=  \prod_{i=1}^\ell \left (1 +  {\rm e}^{\alpha E_i } \right ) 
\EE
with the strength parameter $\alpha$ chosen to maintain the trait average at the fitness optimum, 
\EQ
\langle E \rangle = \frac{\partial}{\partial \alpha} \log Z_g = \sum_{i=1}^\ell \frac{{E_i \rm e}^{\alpha E_i}}{1 + {\rm e}^{\alpha E_i}} = E^*.
\EE
The canonical entropy is then given by $S = \alpha \langle E \rangle - \log Z_g = \alpha E^* - \log Z_g$, which leads to the entropy density~(\ref{varsigma}). 

We conclude that the genotypic predictability is always small for complex traits, because an extensive number of genotypes remains compatible even with a strongly constrained trait value. Only after the projection from genotype to phenotype, selection can generate predictability.

\section{Inference of stabilizing selection}
\label{sec:SelInference}

\begin{figure}[t!]
\includegraphics[width=0.445\textwidth]{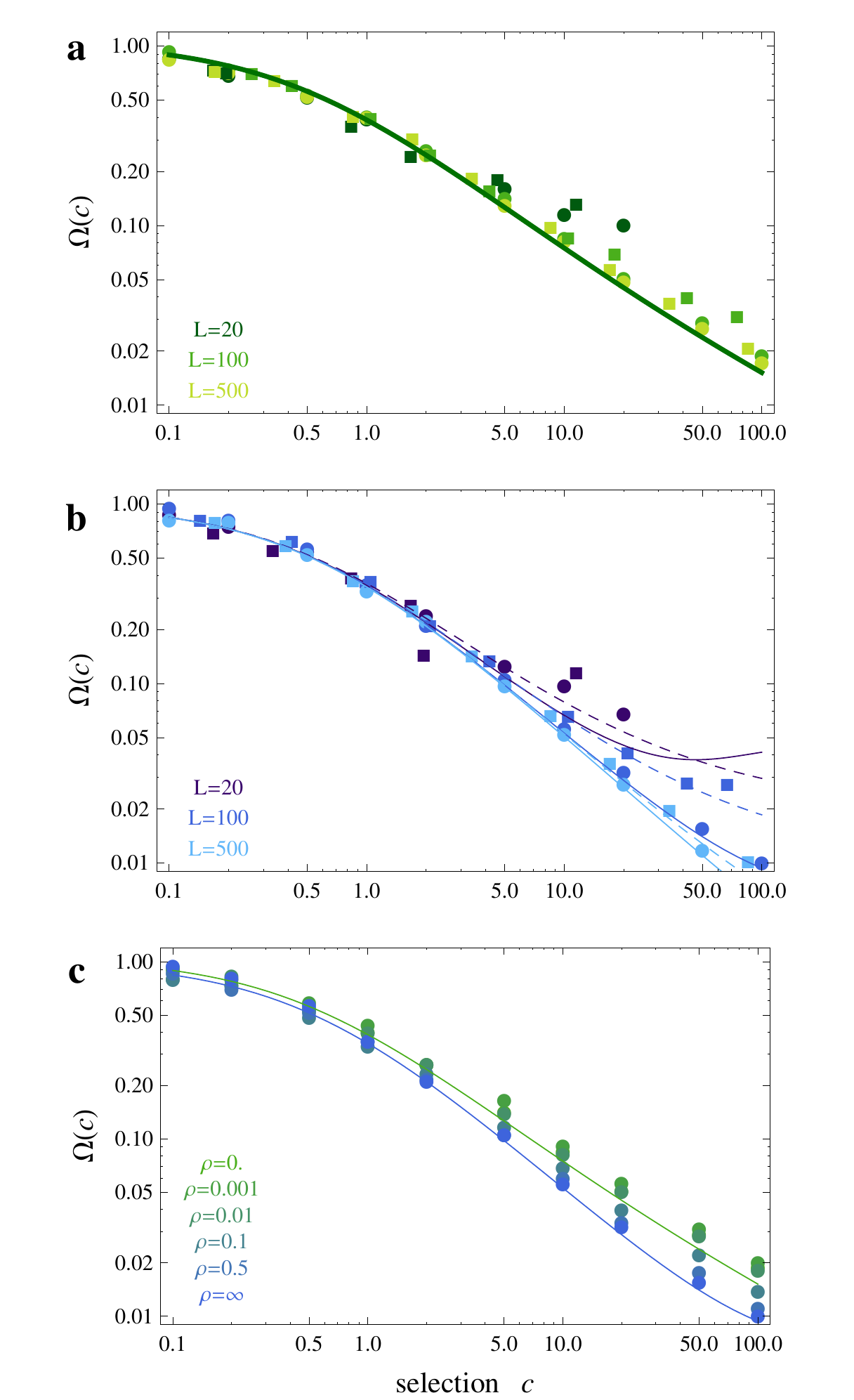}
\caption{
{\bf \small Inference of stabilizing selection.} \small 
The phenotypic observable $\Omega$ measures the ratio between divergence and diversity of a quantitative trait, as given by eq.~(\ref{Omega}). This ratio is plotted against the strength of stabilizing selection, $c$, for populations with different numbers ($\ell$) and effect distributions ($\kappa$) of the trait's constituent sites, and with different recombination rates. 
(a)~Data for non-recombing populations with $\ell = 20, 100, 200$ (dark to light green symbols) and two different effect distributions: delta distribution (all sites have equal effect, circles), exponential distribution (squares). Other system parameters as in Fig.~3. These data are in good agreement with the universal theoretical behavior $\Omega ( c )$ (solid line) given by eq.~(\ref{eq_Omega}). Data points are shown within the range of applicability of the theory, $c /\ell < 1$ (for larger values of $c$, selection becomes balancing for individual loci). 
(b)~Data for populations with free recombination for the same values of $\ell$ (dark to light blue symbols) and the same effect distributions. These data are in good agreement with the theoretical behavior $\Omega ( c )$ (lines) given by eq.~(\ref{eq_Omega}), which contains a small dependence on $\ell$ (dark to blue lines) and on the effect distribution (delta: solid lines, exponential: dashed lines). 
(c)~Data for populations with different recombination rates $\rho = 0.001, 0.01, 0.1, 0.5, \infty$ (blue to green circles), evaluated for $\ell = 100$ and exponential effect distribution. These data interpolate between the theoretical predictions without recombination (green line) and with free recombination (blue line). Together, this shows the nearly universal dependence of the divergence-diversity ratio on the strength of stabilizing selection. 
}
\end{figure}

Our results suggest a method to infer selection on a quantitative trait. The method is based on trait measurements within and across populations, but it does not require knowledge of the trait's genomic basis. Specifically, the ratio 
\EQ
\Omega = 
4\theta \, \frac{\langle (\Gamma - \langle \Gamma \rangle)^2 \rangle}{\langle \Delta \rangle} = 
2\theta \, \frac{\langle (\Gamma_1 - \Gamma_2)^2 \rangle}{\langle \Delta \rangle}
\label{Omega}
\EE
depends only on phenotypic observables: it can be evaluated from the average trait diversity within populations, $\langle \Delta \rangle$, and the variance of the trait mean across populations, $\langle (\Gamma - \langle \Gamma \rangle)^2 \rangle$, at evolutionary equilibrium (we assume the neutral sequence diversity $\theta$ to be known independently). The ensemble variance $\langle (\Gamma - \langle \Gamma \rangle)^2 \rangle$ is just half of the equilibrium divergence, $\langle (\Gamma_1 - \Gamma_2)^2 \rangle$, which, in turn, is close to the divergence between evolutionarily related populations, $\langle (\Gamma(t_1)  - \Gamma (t_2))^2 \rangle$, provided their divergence time is larger than the relaxation time of the trait to equilibrium. This is a reasonable approximation for traits under substantial selection, and our model can be extended to divergence data between closely related populations~\cite{HeldNourmohammadLassig12}. 

Our theory provides an analytical expression for $\Omega$, which is given by eq.~(\ref{eq_Omega}). It shows that $\Omega$ is a monotonically decreasing function of the strength of selection, $c$. This dependence can be used to infer $c$, which is defined as the fitness drop per $2N$ generations at a distance of one neutral standard deviation from the trait optimum. Both $\Omega$ and $c$ are pure numbers, which are independent of the units of trait and fitness. As shown by eq.~(\ref{eq_Omega}), our phenotype-based method is formally similar to the well-known McDonald-Kreitman test, which evaluates divergence and diversity of genomic sequences~\cite{Kreitman:1991vh}. However, the McDonald-Kreitman test has a different scope, which is to infer positive selection. 

The $\Omega$ test exploits a universal characteristic of stabilizing selection: it affects the trait diversity less than its mean. This characteristic is quite intuitive from Fig.~1, which suggests that selection acts on divergence and on diversity with different  characteristic strength. This strength is given by the curvature of the fitness landscape, $c_0$, multiplied with a relevant squared trait scale at neutrality. The basic such scale is the neutral  expectation value of the trait divergence, $\langle \hat \Gamma_2 \rangle_0 \approx E_0^2$. The trait scales within a population are different: Without recombination, selection acts on genotypes, and the relevant scale is the total trait diversity, $\theta E_0^2$. With strong recombination, selection acts on individual trait loci, and the relevant scale is the squared trait amplitude of one such locus, which is of order $E_0^2 / \ell$. Both within-population scales are small against the divergence scale $E_0^2$. 

Most importantly, the inference of selection is confounded neither by number $\ell$ and effect distribution $\kappa$ of the trait's constituent sites, nor by recombination between these sites. All of these genetic factors affect $\Omega$ only through the term ${\cal G}$ in eq.~(\ref{eq_Omega}), which is small in the relevant range of $\theta$ (at most percent) and $\ell$ (at least tens of sites). As a result, $\Omega$ depends on the strength of selection in a nearly universal way. Numerical simulations of populations with different site numbers, effect distributions, and recombination rates confirm this behavior, as shown in Fig.~6.

\section{Discussion}

In this paper, we have developed a statistical model for the evolution of complex molecular traits. We have shown that the dynamics of such traits can be described by approximate Kimura diffusion equations. In an arbitrary fitness landscape, this dynamics leads to coupled evolutionary equilibria for trait mean and diversity. Unlike the standard low-mutation or high-recombination approximations, our model is applicable to correlated multi-site processes, which evolve large linkage disequilibria between the trait's constitutive sites. Such processes govern the evolution of complex traits in asexual populations; in sexual populations, they are relevant for mesoscopic traits, which are polymorphic and based on a genomic region with limited recombination. Our model is a starting point for the analysis of such traits beyond the infinite-recombination assumption of quantitative genetics. It can and should be extended in a number of directions, which include the crossover from genotype selection to allele selection for finite recombination rates~\cite{shraiman:2012}, traits with a nonlinear dependence on genotype, more rugged fitness landscapes, and time-dependent fitness ``seascapes'' driving adaptive trait evolution~\cite{Mustonen:2009vu}. 

Our model leads to a new, quantitative test for stabilizing selection on quantitative traits, which is based on the ratio between trait divergence and trait diversity at equilibrium. We have shown that this ratio measures the strength of stabilizing selection in a nearly universal way, independently of the trait's genomic basis and of the recombination rate. This test can also be extended to quantitative traits  in a time-dependent fitness seascape, which will be the subject of a forthcoming companion paper~\cite{HeldNourmohammadLassig12}. 

Complex phenotypes integrate the information of multiple genomic sites. Compared to their constitutive genotypes, they represent biological functions on a larger scale. Both at the genomic and at the phenotypic level, we can ask about the predictability of evolution: How informative is sequencing or trait measurements in one population about the same quantities in a different population that evolves in the same fitness landscape? As we have shown in section~\ref{predictability}, this question can be made precise by defining predictability in terms of an entropy difference between intra- and cross-population distributions of genotypes or trait values. For complex traits, predictability turns out to depend on scale and on selection. There is little predictability at the genome level, because the total number of genotypes encoding a functional trait is vastly larger than that realized in any one population. The equilibrium predictability is exponentially small in the number of trait sites, and populations evolving from a common ancestor will diverge through mutations at different sites. At the phenotypic level, the equilibrium predictability is related to the divergence-diversity ratio $\Omega$, as given by eq.~(\ref{calP_eq}). It is small at neutrality, but  under sufficiently strong stabilizing selection, it can reach values of order one. Hence, stabilizing selection generates predictability of evolution at the phenotypic level.

\newpage{}

\bibliography{citemin}

\newpage{}

\section*{Appendix: \\
Non-equilibrium ensembles of quantitative traits}

Here we analyze the evolution equation (\ref{PGDt}) for the joint distribution $Q(\Gamma, \Delta,t)$ of trait mean and variance in asexual populations. We consider the case of stabilizing selection in the fitness landscape $F(\Gamma, \Delta)$ given by eq.~(\ref{FGammaDeltastabsel}). Using the scaled trait variables $\hat\gamma=(\Gamma-\Gamma_0)/\Tr_0$, $\delta=\Delta/\Tr_0^2$ and $\lambda=(\Gamma-\Tr^\star)/\Tr_0$ and the scaled selection strength  $c=2N\Tr_0^2 c_0$ defined in eqs.~(\ref{scaling}) and (\ref{cscaling}), this equation can be written in the form 
\begin{eqnarray}
2N\, \frac{\partial}{\partial t} \hat Q (\hat \gamma, \delta ) & = & 
\frac{\partial}{\partial \hat \gamma} 
\left [ \frac{\partial}{\partial \hat \gamma} \, \delta+4\theta  \hat\gamma+2c \ \delta(\hat\gamma- \langle\lambda\rangle_0) \right ]
Q(\hat\gamma,\delta,t) + 
\nonumber \\
&&
\frac{\partial}{\partial \delta} 
\left [ \frac{\partial}{\partial \delta} \, 2\delta^2+8\theta (\delta-1) +2\delta+2c\ \delta^2]\right ] Q(\hat\gamma,\delta,t) 
\nonumber \\
& = & \frac{\partial}{\partial \hat \gamma} \hat J^{\gamma} (\hat \gamma, \delta, t) + \frac{\partial}{\partial \delta} \hat J^\delta (\hat \gamma, \delta, t), 
\label{QGDt_scal}
\end{eqnarray}
where $(\hat J^{\gamma}, \hat J^\delta)$ denotes the probability current in the $\gamma$-$\delta$ plane. The scaled distribution and current are related to their unscaled counterparts,
\EQA
Q(\Gamma, \Delta; c_0, E_0) & = & \Tr_0^{-3/2} \, \hat Q(\hat{\gamma},\delta; c),
\label{Qhat}
\\
J^\Gamma (\Gamma, \Delta; c_0, E_0) & = & E_0^{-1} \, \hat J^\gamma (\hat \gamma, \delta; c),
\nonumber \\
J^\Delta (\Gamma, \Delta; c_0, E_0) & = & E_0^{-1/2} \hat  J^\delta (\hat \gamma, \delta; c)
\label{Jhat}
\EEA
If we keep the trait effect distribution $\kappa(\epsilon)$ of individual constituent sites fixed, the squared overall trait scale is proportional to their number, $E_0^2=\kappa_2\ell/4$. Hence, we can interpret the relations (\ref{Qhat}), (\ref{Jhat}) as scale transformations relating systems with different values of $\ell$. 

The joint dynamics of $\Gamma$ and $\Delta$ does not have an equilibrium solution, because the mutation coefficient field $(4 \theta \hat \gamma, 8\theta (\delta-1) +2\delta)$ is non-integrable. As shown by numerical simulations, this dynamics leads instead to a non-equilibrium stationary distribution $\hat Q_{\rm stat} (\hat \gamma, \delta)$ (shown in Fig.~2),  which has with a finite current $(\hat J^\gamma, \hat J^\delta) (\hat \gamma, \delta)$. According to eqs.~(\ref{Qhat}) and (\ref{Jhat}), we obtain a scale-invariant stationary non-equilibrium state\footnote{The scale-invariant state (\ref{Qhat}, \ref{Jhat}) of asexual neutral evolution ($c = 0$) can be broken by two relevant perturbations: recombination (which changes the scaling of diversity fluctuations from $\Delta \sim \ell$ to $\hat \Delta \sim \ell^{1/2}$) and stabilizing selection of constant unscaled strength $c_0$ (which generates  a trait autocorrelation $\langle C \rangle < 0$). }
describing a family of systems with different $\ell$ and constant scaled selection strength $c$. In this family of systems, a finite stationary current persists for large values of $\ell$. 

To test this prediction, we measure the current by binning changes of $\Gamma$ and $\Delta$ through discretized grid lines in our simulation. At each junction in the grid, we
record positive and negative changes separately. These changes determine the current components $J^\Gamma (\Gamma, \Delta) = J_+^\Gamma (\Gamma, \Delta) - J_-^\Gamma (\Gamma, \Delta)$ and  $J^\Delta (\Gamma, \Delta) = J_+^\Delta (\Gamma, \Delta) - J_-^\Delta (\Gamma, \Delta)$, and we obtain dimensionless measures for the violation of detailed balance,
\EQ
{\cal J}_\delta(\gamma) \equiv 
\frac{J_+^\Delta (\Gamma) - J_-^\Delta (\Gamma)}{
J_+^\Delta (\Gamma) + J_-^\Delta (\Gamma)}, 
\hspace{1cm}
{\cal J}_\gamma (\delta) \equiv 
\frac{J_+^\Gamma (\Delta) - J_-^\Gamma (\Delta)}{
J_+^\Gamma (\Delta) + J_-^\Gamma (\Delta)} ,
\label{calJ}
\EE
which are obtained from the integrated currents 
$J_\pm^\Delta (\Gamma) \equiv \int J_\pm^\Delta (\Gamma, \Delta) \, d \Delta$ and
$J_\pm^\Gamma (\Delta) \equiv \int J_\pm^\Gamma (\Gamma, \Delta) \, d \Gamma$ (these measures are less noisy than their local counterparts). In Fig.~\ref{fig:current}(a), we show the ratios (\ref{calJ}) for systems with different values of $\ell$ (at a fixed value of $c$). As predicted by the scale transformation (\ref{Jhat}), these data collapse onto unique functions ${\cal J}_{\delta} (\gamma)$ and ${\cal J}_\gamma (\delta)$. They reveal a closed stationary current with a clockwise loop for $\hat \gamma > 0$ and a symmetrical, counterclockwise loop for $\hat \gamma < 0$.

\begin{figure}
\includegraphics[width=\textwidth]{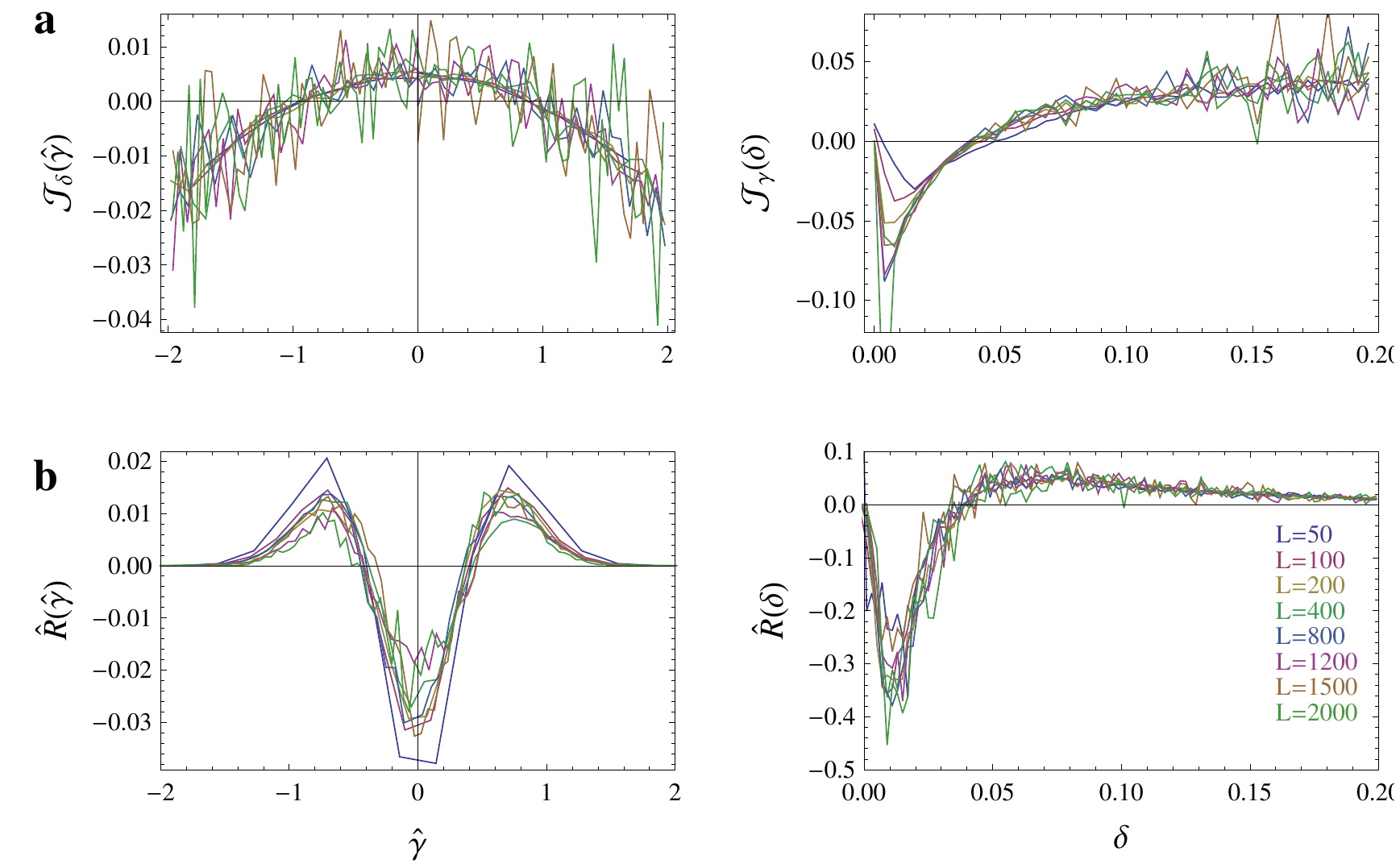}
\caption{\label{fig:current}
{\bf \small Deviations from equilibrium under complete genetic linkage.} \small
(a)~Current ratios ${\cal J}_\delta(\gamma)$ (left panel) and ${\cal J}_\gamma (\delta)$ (right panel), as defined in eq.~(\ref{calJ}).  
(b)~Deviations $\hat R(\hat\gamma)=E_0 R(\Gamma)$ (left panel) and $\hat R(\delta)=E_0^2R(\Delta)$ (right panel) of the stationary distribution from the Boltzmann form, as defined in eq.~(\ref{R}). These data are shown for systems with $\ell = 100, 120, \dots, 200$ constitutive sites of equal effect.   They collapse into unique functions, indicating a scale-invariant non-equilibrium state with stationary current; see eqs.~(\ref{Qhat}) and (\ref{Jhat}). The fitness optimum is set to $E^\star=\Gamma_0=\ell/2$. Other system parameters are as in Fig.~\ref{fig:traitdist}. 
}
\end{figure}

The breakdown of detailed balance in the stationary state has an important consequence: the distribution $Q_{\rm stat} (\Gamma, \Delta)$ under selection is no longer of Boltzmann form. The deviations are measured by the function 
\EQ
R (\Gamma, \Delta) \equiv Q_{\rm stat} (\Gamma, \Delta) - \frac{1}{Z} \, Q_0 (\Gamma, \Delta) \exp[2 N F(\Gamma, \Delta)]. 
\label{R}
\EE
Fig.~\ref{fig:current}(b) shows the marginal differences $R(\Gamma) \equiv \int R(\Gamma, \Delta) d \Delta$ and $R(\Delta) \equiv \int R(\Gamma, \Delta) d \Gamma$ for systems with different values of $\ell$ (at a fixed value of $c$). After rescaling according to eq.~(\ref{Qhat}), these data again collapse onto a single function $\hat R (\hat \gamma, \delta)$. The actual stationary distribution $ Q_{\rm stat} (\Gamma, \Delta)$ is seen to be broader than the corresponding Boltzmann distribution.

\end{document}